\documentclass[11pt, oneside]{article}   	
\usepackage{jheppub}
\usepackage{amsmath}
\usepackage{amssymb}
\usepackage{amsfonts}
\usepackage{amsthm}
\usepackage{amsbsy}
\usepackage{array}
\usepackage{mathtools}
\usepackage[usenames,dvipsnames]{xcolor}
\usepackage{subcaption}
\usepackage{dsdshorthand}
\usepackage{graphicx}
\usepackage[vcentermath]{youngtab}

\makeatletter
\def\@fpheader{\ }
\makeatother

\title{Snowmass White Paper: \\The Numerical Conformal Bootstrap}
\author[a]{David Poland,}
\author[b]{David Simmons-Duffin}
\affiliation[a]{Department of Physics, Yale University, New Haven, CT 06520, USA}
\affiliation[b]{Walter Burke Institute for Theoretical Physics, Caltech, Pasadena, California 91125, USA}

\date{}
\abstract{We give a brief overview of the status of the numerical conformal bootstrap.}

\preprint{CALT-TH 2022-013}

\begin{document}

\maketitle
\pagenumbering{roman}
\setcounter{page}{2}
\newpage
\pagenumbering{arabic}
\setcounter{page}{1}

\section{Introduction}

The idea of the {\it conformal bootstrap\/} is to constrain and solve CFTs using physical consistency conditions like symmetry, unitarity, and causality. By relying on nonperturbative structures, bootstrap methods can work even in strongly-coupled systems where traditional perturbative techniques fail. Over the last few decades, the conformal bootstrap idea has crystalized into two concrete strategies: 1) exploiting exact solvability and 2) deriving bounds from positivity.  An early success of the first strategy was the exact solution by Belavin, Polyakov, and Zamolodchikov (BPZ) of an infinite class of 2d CFTs, obtained by identifying a powerful infinite-dimensional symmetry group in these theories \cite{Belavin:1984vu}. The second strategy, pioneered by Rattazzi, Rychkov, Tonni, and Vichi in \cite{Rattazzi:2008pe} proceeds by formulating sum rules with positivity properties and using convex optimization to extract information from the sum rules. A virtue of this second strategy is its flexibility: nontrivial positivity bounds can be derived for theories with a wide variety of symmetry groups and consistency conditions, even when exact solvability is impossible. When combined with numerical convex optimization techniques, this approach is called the {\it numerical conformal bootstrap}.

In recent years, the numerical conformal bootstrap has led to a treasure-trove of new results, including precise determinations of critical exponents in physically-relevant theories such as the 3d Ising and $O(N)$ models,  constraints on theoretically important theories such as 4d $\mathcal{N} = 4$ supersymmetric Yang-Mills theory and the 6d $\cN=(2,0)$ theory, and has inspired promising new ideas for bootstrapping S-matrices, modular-invariant partition functions, and more. Many of these developments were summarized in the recent review article~\cite{Poland:2018epd}. However, the numerical bootstrap is a rapidly-growing field and we anticipate that it will play a central role in theoretical physics over the next decade. In this white paper we give an (incomplete) overview of some of the targets for the numerical conformal bootstrap program, focusing on CFTs in 3 and 4 dimensions, along with a summary of current state-of-the-art algorithms, software tools, current and future computational needs, and open questions. Closely related Snowmass white papers are those on the analytic bootstrap~\cite{Hartman:2022zik}, bootstrapping string theory~\cite{Gopakumar:2022kof}, the S-matrix bootstrap~\cite{Kruczenski:2022lot}, superconformal field theories~\cite{Argyres:2022mnu}, and Hamiltonian truncation~\cite{Fitzpatrick:2022dwq}.

\section{Targets}

In addition to theoretical beauty, BPZ's solution of rational 2d CFTs has practical significance: Such theories describe critical points in numerous condensed matter and statistical systems, both in theoretical settings and in the laboratory. An experimentalist studying a 2d critical point can measure a few exponents and then compare to a catalog of known solved 2d CFTs to learn a lot about their system. Numerical bootstrap methods offer hope for similarly useful solutions of higher-dimensional CFTs. From this point of view, 3d and 4d CFTs with a small number of degrees of freedom are particularly interesting targets for numerical bootstrap methods. Below, we list some of these targets, recounting what has been computed so far and highlighting interesting future directions.

\subsection{3d Scalar Models}

One of the simplest classes of targets for the conformal bootstrap consist of fixed points that can be described by interactions between one or more scalar fields. The $O(N)$ vector models (labeled by a positive integer $N$) are a family of conformal field theories in 3 spacetime dimensions that describe second-order phase transitions in numerous condensed matter and statistical systems. The models can be arrived at through an RG flow from a UV Lagrangian with $\mathbb{Z}_2$ or $O(N)$ invariant interactions $\mathcal{L}_{int} \sim \lambda (\phi_i \phi_i)^2$, where $\phi_i$ is an $N$-component scalar field.  A comprehensive review of the known CFT data in the $O(N)$ models can be found in~\cite{Henriksson:2022rnm}. 

The $N=1$ model, also called the 3d Ising CFT, describes liquid-vapor critical points, binary fluid mixtures, uni-axial magnets, and more. This theory has been a focal point for the conformal bootstrap since the works~\cite{El-Showk:2012cjh,El-Showk:2013nia,El-Showk:2014dwa}, which established the existence of a ``kink" at the 3d Ising scaling dimensions in the general bound on the leading $\mathbb{Z}_2$-even scaling dimension $\Delta_{\epsilon} = 3 - \frac{1}{\nu}$ as a function of the leading $\mathbb{Z}_2$-odd scaling dimension $\Delta_{\sigma} = \frac12 + \frac{\eta}{2}$. Subsequent work showed that their scaling dimensions and OPE coefficients could be robustly isolated into closed regions of parameter space by considering bootstrap constraints from mixed correlators containing both $\s$ and $\e$, after imposing the condition that $\s$ and $\e$ are the only relevant operators in the theory~\cite{Kos:2014bka,Kos:2016ysd}.\footnote{Related progress was also made at constraining the long-range Ising model using bootstrap methods~\cite{Behan:2017dwr,Behan:2017emf,Behan:2018hfx}. In general there are many non-local CFTs which could be obtained by taking a local CFT and coupling one of its operators to a generalized free field. Depending on one's perspective these are either interesting targets for the bootstrap in their own right, or annoyances that one must figure out how to exclude in order to hone in on the local CFTs of interest.}

The 3d Ising CFT has also served as an important benchmark for comparing analytical bootstrap methods with numerical estimates of the leading-twist trajectories from the extremal functional approach~\cite{Simmons-Duffin:2016wlq,Albayrak:2019gnz,Caron-Huot:2020ouj}. Most recently it has been explored using ``navigator function" methods to constrain irrelevant operators~\cite{Reehorst:2021hmp} and to develop hybrid numerical-analytical approaches to the bootstrap~\cite{Su:2022xnj}. 

The 3d $O(2)$ model is distinguished for both its scientific interest and for its position at the heart of a controversy. As the theory describing the superfluid transition in liquid helium (as well as thin-film superconductors \cite{PhysRevB.44.6883,PhysRevLett.95.180603}), the $O(2)$ model is the cleanest known critical system for experiment. The best measurements of the $O(2)$ model (indeed, the current most precise measurements of any critical system) were performed as part of the Lambda-Point experiment on the space shuttle in the early 90's \cite{PhysRevB.68.174518}. Unfortunately, the results of that experiment and the current best Monte-Carlo simulations \cite{PhysRevB.74.144506} {\it disagreed} with each other with $8\sigma$ significance, indicating a severe issue with either the experiment or simulations. The numerical bootstrap has since produced precise bounds on the scaling dimensions and OPE coefficients of this model~\cite{Go:2019lke, Chester:2019ifh}, giving results in remarkable agreement with those from Monte-Carlo simulations. 

The 3d $O(3)$ model describes Heisenberg magnets, which are systems where the magnetization can point in any direction in 3-dimensional space. Na\"ively, such magnets can arise in any material that is sufficiently isotropic that $O(3)$ symmetry can emerge at long distances. However, whether $O(3)$ symmetry can emerge in practice is not obvious. In lattices with cubic symmetry, $O(3)$-symmetry breaking properties of the lattice might survive at long distances, making it impossible to realize a true Heisenberg magnet in these systems. Whether $O(3)$ symmetry breaks or not is a several-decades-old open question that has been attacked using a variety of non-rigorous theoretical techniques \cite{doi:10.1080/00150198908245184,PhysRevB.61.14660,PhysRevB.61.15130,Caselle_1998,PhysRevE.94.042107,PhysRevB.62.12195,PhysRevB.27.609,PhysRevLett.33.813,SHALAEV1997105,KLEINERT1995284,PhysRevB.56.14428,PhysRevB.65.144520,PhysRevB.61.15136,Pakhnin2002}. 

The question of whether $O(3)$ symmetry is emergent in cubic lattice models can be translated into the language of conformal field theory: it is equivalent to the question of whether the leading $O(3)$ four-index tensor operator $\cO_{ijkl}$ in the $O(3)$ model is relevant (has dimension $<3$) or not. This question can be addressed using the conformal bootstrap: by studying a system of correlation functions analogous to the ones studied for the $O(2)$ model, one can compute upper bounds on the dimension of $\cO_{ijkl}$ and determine whether it is relevant. In~\cite{Chester:2020iyt} a precision study of the $O(3)$ model was performed using the state-of-the-art bootstrap methods, computing its critical exponents to high precision and resolving this old question, demonstrating that the four-index tensor is relevant and should trigger a flow to the fixed point with cubic symmetry.

Bootstrapping the cubic model itself was initiated in~\cite{Rong:2017cow,Stergiou:2018gjj}. The work \cite{Stergiou:2018gjj}~also found evidence for a new 3d CFT with cubic symmetry called a ``Platonic CFT", possibly relevant for structural phase transitions. This model was studied further using mixed correlators in~\cite{Kousvos:2018rhl,Kousvos:2019hgc}. More work must be done to isolate the cubic fixed point relevant for Heisenberg magnets using bootstrap methods and to clarify the nature of the ``Platonic CFT". A closely-related fixed point is the ``biconical" fixed point with $O(2) \times \mathbb{Z}_2 \subset O(3)$ symmetry, which is also a natural target for the bootstrap. More intricate scalar models on which progress has been made are those with $O(N) \times O(M)$ symmetry~\cite{Nakayama:2014lva,Nakayama:2014sba,Henriksson:2020fqi}, hypercubic models~\cite{Stergiou:2018gjj}, projective models~\cite{Reehorst:2020phk}, and MN models~\cite{Stergiou:2019dcv,Henriksson:2021lwn,Kousvos:2021rar}. 

Finally, exciting progress has been made at implementing the 3d bootstrap with external global symmetry currents~\cite{Dymarsky:2017xzb, Reehorst:2019pzi} and stress tensors~\cite{Dymarsky:2017yzx}. These studies have been able to place interesting constraints on 3-point functions involving conserved operators in the Ising and O(2) models, as well as produce interesting general maps of the allowed space of operator dimensions. An important future direction will be to improve these studies by exploring constraints from mixed correlators involving stress tensors and scalars, non-Abelian currents, and other operators. Via the light-ray OPE \cite{Hofman:2008ar,Kologlu:2019mfz,Dixon:2019uzg,Chang:2020qpj}, the OPE data of stress tensors and symmetry currents plays an important role in event shapes such as energy and charge correlators. By computing this data in specific theories, one could make predictions for event shapes in strongly-coupled critical systems, which could perhaps make contact with future experiments.

\subsection{3d Fermionic Models}
Another interesting class of models are 3d CFTs containing $N$ Majorana fermions $\psi_i$ with Yukawa couplings to one or more scalar fields, often called Gross-Neveu-Yukawa (GNY) models. In the simplest model one has $\mathcal{L}_{int} \sim g \phi \bar{\psi}_i \psi_i + \lambda \phi^4$, in which case the model has an $O(N)$ symmetry. Another variant breaks the fermions into two groups with interactions of opposite signs $\mathcal{L}_{int} \sim g \phi (\bar{\psi}^+_i \psi^+_i - \bar{\psi}^-_i \psi^-_i) + \lambda \phi^4$, in which case there is an $O(N/2) \times O(N/2) \rtimes \mathbb{Z}_2$ symmetry. Other important models include the ``Chiral XY" and "Chiral Heisenberg" GNY models involving two or three scalar fields. These models have various proposed applications in condensed matter systems, including quantum phase transitions in graphene~\cite{herbut2006interactions, herbut2009relativistic, Mihaila:2017ble} and d-wave superconductors~\cite{vojta2000quantum, vojta2003quantum}. The simplest model with $N=1$ Majorana fermion coupled to a real scalar is believed to possess emergent supersymmetry and correspond to the minimal 3d $\mathcal{N} = 1$ supersymmetric extension of the Ising model, proposed to have a realization on the boundary of topological superconductors~\cite{Grover:2013rc}. 

Preliminary studies applying the bootstrap to fermion 4-point functions showed the GNY models to be promising targets for the bootstrap~\cite{Iliesiu:2015qra,Iliesiu:2017nrv}. In particular, the resulting bounds possess kinks that match to estimates from large-$N$ and $\epsilon$-expansion methods. An interesting comparison between these methods can be found in~\cite{Ihrig:2018hho}. Ongoing bootstrap work involving mixed correlators further shows that these theories can be numerically isolated into islands similar to those of the $O(N)$ vector models~\cite{GNYtoappear}. Similarly, the 3d $\mathcal{N} = 1$ supersymmetric Ising model has been bootstrapped into a precise island using the constraints of supersymmetry on mixed correlators containing scalar operators~\cite{Rong:2018okz, Atanasov:2018kqw, Atanasov:2022bpi}. Interesting bootstrap constraints have also been placed the 3d $\mathcal{N} = 2$ supersymmetric extension of the Ising model and ~\cite{Bobev:2015vsa, Bobev:2015jxa} as well as on more general Wess-Zumino models~\cite{Baggio:2017mas, Rong:2019qer}.

The 3d fermion bootstrap also revealed an unexplained kink/jump in the bounds on the leading parity even/odd scalar operators, which was conjectured to correspond to a previously unknown ``dead-end" CFT with no relevant scalar perturbations~\cite{Iliesiu:2015qra}. Subsequently it was thought that this feature may be a numerical artifact due to the ``fake primary" effect~\cite{Karateev:2019pvw}, but a recent study showcasing the new \texttt{blocks\_3d} software revisited this possibility and found that the ``fake primary" effect couldn't fully explain the feature~\cite{Erramilli:2020rlr}. Thus, the possibility of a new ``dead-end" 3d CFT remains open and merits further study.

\subsection{3d Gauge Theories}

Many other interesting 3d CFTs can be reached in 3d gauge theories coupled to scalars or fermions. An important target is 3d QED coupled to $N_f$ fermions. These theories have physical relevance for Dirac spin liquids, and also serve as an important testing ground for ideas related to conformal windows and non-supersymmetric dualities. Progress using the bootstrap has been made in a variety of works~\cite{Chester:2016wrc,Chester:2017vdh,Li:2018lyb,Li:2020bnb,He:2020azu,Li:2021emd,He:2021sto,Albayrak:2021xtd}. 

In particular,~\cite{Chester:2016wrc, Chester:2017vdh} initiated a study of the bootstrap applied to 4-point functions of monopole operators and found bounds that are close to being saturated by predictions from the large $N_f$ expansion. The works~\cite{Li:2018lyb,Li:2020bnb,He:2020azu} found an interesting kink structure in bounds from 4-point functions of fermion bilinear operators. These ``non-Wilson-Fisher" kinks in 3d were conjectured to be related to 3d QED in~\cite{Li:2018lyb}; this interpretation suggests a critical $N_f$ between $2$ and $4$. 

The recent works~\cite{He:2021sto, Albayrak:2021xtd} pursued mixed correlators between monopoles and fermion bilinears under various assumptions for the special case $N_f = 4$. Currently there are interesting tensions between large $N_f$ calculations extrapolated to $N_f = 4$, Monte Carlo studies of monopole operators, and the relevance or irrelevance of 4-fermion operators which are needed for lattice realizations of the fixed point. The bootstrap results so far place strong constraints on the allowed parameter space (and are potentially compatible with the large $N_f$ predictions~\cite{Albayrak:2021xtd}), but are not yet fully conclusive.

A related class of theories is 3d QED coupled to $N$ complex scalars (sometimes called the ``non-compact CP${}^{N-1}$" model), which is related to the phenomenon of ``deconfined criticality". In the case of $N=2$, the bootstrap has placed strong constraints on the possibility of the $SU(2)$ or ``easy-plane" fixed points having emergent $SO(5)$ or $O(4)$ symmetry~\cite{Nakayama:2016jhq, DSD:2016, DP:2017, Iliesiu:2018, Poland:2018epd}, contributing evidence that the $N=2$ theories are likely weakly first-order transitions. This is also supported by a more direct bootstrap study targeting $N_f = 2$ fermionic QED~\cite{Li:2021emd} (conjecturally related to the ``easy-plane" $N=2$ scalar QED by duality), which also shows that the bootstrap bounds are not easily compatible with Monte Carlo data and a first-order transition is the most likely scenario.

At larger values of $N$, some promising bootstrap progress has been made by considering scalar adjoint bilinears. In particular,~\cite{Manenti:2021elk} found intriguing kinks in their bounds and~\cite{He:2021xvg} found isolated islands at large $N$ and in $2+\epsilon$ dimensions, after inputting gaps motivated by equations of motion. However, the 3d scalar QED theories still remain to be isolated using bootstrap methods at small $N$, and this remains an exciting target for the bootstrap.

Many more complicated variations of these non-supersymmetric models exist, e.g.~3d QED coupled to both fermions and scalars, 3d QCD coupled to matter in various representations, 3d Chern-Simons theories coupled to matter, and so on. It will be an important task for the future of the bootstrap to learn how to disentangle and isolate each these fixed points from each other.

Finally, let us highlight that exciting progress has been made in applying the numerical bootstrap to $\mathcal{N} = 6,8$ supersymmetric ABJ(M) theories. In particular very precise islands for OPE coefficients of protected operators have now been computed in the works~\cite{Agmon:2017xes,Agmon:2019imm,Binder:2020ckj,Alday:2021ymb}. Some of these results have been highlighted in the Snowmass white paper on bootstrapping string theory~\cite{Gopakumar:2022kof}. With additional work, an improved understanding of the relevant superconformal blocks, and the incorporation of information from localization, it is likely that the numerical bootstrap can extend these precise results to a much broader class of observables in these theories.

\subsection{4d Gauge Theories}

4d non-Abelian gauge theories are believed to exhibit fixed-points with a wide variety of gauge groups and matter contents. Given a particular gauge group, the set of allowed matter content that leads to a nontrivial interacting IR fixed-point is called the {\it conformal window}. The conformal window is known precisely in supersymmetric 4d gauge theories, but characterizing the conformal window of nonsupersymmetric 4d gauge theories is a longstanding open problem. This seems like a natural problem for bootstrap approaches, but it has so far proved technically challenging. One reason is that bootstrap methods are most powerful when supplemented with information that can isolate a theory from other CFTs. (For example, to isolate the 3d Ising model, it is helpful to input that the theory has a $\Z_2$ symmetry and only two relevant operators.) However, information about the gauge group of non-Abelian gauge theories is encoded in ways that are hard for current bootstrap methods to access. For example, the gauge group affects the spectrum of baryons, but these are likely relatively high dimension operators, which generally exhibit weaker bootstrap bounds. An important challenge is to find observables that are amenable to a bootstrap analysis, and that provide simple access to the defining data of non-Abelian gauge theories.

Instead of attempting to isolate individual conformal gauge theories, most bootstrap studies so far have focused on computing universal bounds on classes of theories.
For example, general lower bounds on the anomalous dimension of the fermion bi-fundamental $\bar{\psi}_L \psi_{R}$ can be computed using the bootstrap, assuming irrelevance of the leading singlet, giving relatively weak bounds like $\Delta_{\bar{\psi} \psi} > 1.21$ for $N_f = 8$~\cite{Nakayama:2016knq}. Similar bounds assuming staggered fermion realizations are possible on the lattice were computed in~\cite{Iha:2016ppj}. 

More recently, intriguing kinks/discontinuities were noticed in 4d bounds on the dimensions of 4-fermion operators~\cite{Li:2020bnb,He:2020azu}, which approach the free fermion value $\Delta_{\bar{\psi} \psi} \sim 3$ from below at large $N_f$. For the commonly-studied case of $N_f = 12$, the kink location yields a fermion bilinear dimension $\sim 2.78$, intriguingly close to lattice estimates for 12 flavor QCD. However, the identified bootstrap solution disagrees with QCD in other ways, e.g.~the leading singlet has a dimension which is much too high. More work needs to be done to firmly establish whether there is a connection between these 4d kinks and the conformal window of 4d QCD.

In order to make further progress on these theories, it is plausible that bootstrap constraints from 4-point functions of spinning operators such as global symmetry currents and the stress tensor need to be incorporated (giving access to information about 't Hooft anomalies). It may also be helpful to incorporate constraints from baryon operators. Some substantial progress at implementing the bootstrap for 4d fermions was made in~\cite{Karateev:2019pvw}, but the bootstrap for currents and stress tensors has not yet been carried out in 4d.

A variety of work has been done on 4d superconformal field theories using the bootstrap; here we will only give some highlights. A primary target for the 4d $\mathcal{N}=1$ superconformal bootstrap is the conformal window of SQCD, but these theories have not yet been isolated using bootstrap methods. Bounds on OPE coefficients related to 't Hooft anomalies (which can be exactly computed) were computed in various works and compared with SQCD~\cite{Poland:2010wg,Poland:2011ey,Berkooz:2014yda,Li:2017ddj,Lin:2019vgi}, but they are not yet close to being saturated. Another intriguing target is a mysterious kink which appears in the bootstrap bound on the leading non-chiral operator appearing in the chiral-antichiral OPE~\cite{Poland:2011ey}. This kink has been conjectured to correspond to a ``minimal" 4d SCFT and its properties were studied in some detail in~\cite{Poland:2015mta,Li:2017ddj}. However, whether this corresponds to a real SCFT, and how to properly isolate it using bootstrap methods, remains uncertain.

Impressive bootstrap bounds have also been computed assuming 4d $\mathcal{N} = 2$ superconformal symmetry, beginning with~\cite{Beem:2014zpa}. In particular the Argyres-Douglas fixed points seem to be excellent targets for the conformal bootstrap, and strong constraints on the OPE coefficients of the simplest Argyres-Douglas fixed points were placed in~\cite{Lemos:2015awa,Cornagliotto:2017snu,Gimenez-Grau:2020jrx,Bissi:2021rei}. Further bootstrap studies of Argyres-Douglas fixed points, as well as of 4d $\mathcal{N} = 2$ SQCD and similar supersymmetric gauge theories, are almost certain to be fruitful arenas for the numerical bootstrap.

Finally, let us highlight the exciting and impressive progress that has been made in bootstrapping 4d $\mathcal{N}=4$ supersymmetric Yang-Mills theory, initiated in the works~\cite{Beem:2013qxa, Alday:2013opa, Beem:2016wfs}. (Similar progress has also been made on the 6d (2,0) SCFTs~\cite{Beem:2015aoa, Lemos:2021azv}.) These works considered 4-point functions of the $20'$ operator and computed a variety of bounds on operator dimensions and OPE coefficients as a function of the central charge, conjecturing that these bounds were saturated by one of the self-dual values of the coupling $\tau = i$ or $\tau = e^{i \pi/3}$. These bounds were generalized to mixed correlators in~\cite{Bissi:2020jve} and compared to new large $c$ calculations in~\cite{Alday:2021vfb}, showing that the numerical bounds were seemingly not saturated at either self-dual value of the coupling. 

More recently, however, the bounds for $SU(2)$ and $SU(3)$ gauge groups were strengthened in~\cite{Chester:2021aun} by including localization information, so that they could be computed as a function of the coupling, showing that the stronger dimension bounds are in fact maximized at $\tau = e^{i \pi/3}$. This method allows for precise bootstrap computations of the Konishi operator dimension as a function of coupling, along with many other results. There will be many more exciting results coming from combining localization computations with the superconformal bootstrap, as this method can be generalized to many other theories.

\subsection{Defects and Boundaries}

Another fruitful direction for the numerical bootstrap has been to consider CFTs in the presence of defects, boundaries, and interfaces. For example, local operators living on the $\mathbb{Z}_2$ twist line defect of the 3d Ising model were studied in~\cite{Gaiotto:2013nva,Ghosh:2021ruh}, while bootstrap constraints on the 3d Ising and $O(N)$ models in the presence of a boundary were studied in~\cite{Liendo:2012hy, Gliozzi:2015qsa, Gliozzi:2016cmg, Padayasi:2021sik}. Here there are different known choices of boundary conditions corresponding to ``ordinary", ``special", and ``extraordinary" transitions. In particular, let us highlight the recent work~\cite{Padayasi:2021sik} which derived strong bootstrap constraints on the extraordinary transitions using semi-definite programming methods (after making some positivity assumptions), which are nicely compatible with Monte Carlo results. 

More generally, one can ask if there is a classification of possible conformal boundary conditions that a given bulk CFT could have. In general this is a hard problem, but some interesting progress was recently made in~\cite{Behan:2020nsf,Behan:2021tcn} using numerical bootstrap methods, giving evidence for possible new boundary conditions that could be satisfied by bulk scalar fields. The problem is more tractable in 2d, where extra symmetries and modular invariance lead to powerful constraints, see the recent work \cite{Collier:2021ngi}. In the future it will be exciting to further combine together the constraints from the bulk and defect bootstrap setups to enable further progress on our understanding of conformal boundary conditions.

Significant progress has also been made in applying numerical bootstrap methods to supersymmetric defects, e.g. to line defects in 4d $\mathcal{N}=4$ SCFTs~\cite{Liendo:2016ymz,Liendo:2018ukf} and 4d $\mathcal{N}=2$ SCFTs~\cite{Gimenez-Grau:2019hez}. This is in addition to immense progress that has been made using analytical bootstrap methods~\cite{Hartman:2022zik}. An interesting hybrid approach combining the numerical bootstrap with information from integrability was also developed recently in~\cite{Cavaglia:2021bnz,Cavaglia:2022qpg}. In the future it is likely that much more can be learned about defect CFTs by generalizing these analyses to mixed correlator systems, and by improving numerical bootstrap approaches in systems where the crossing relations lack positivity.

\section{Algorithms and Software Tools}

It is clear from past and present bootstrap studies that a tremendous amount of information is encoded in the crossing symmetry equations of CFTs. However, to date only a small portion of this information has been used --- for example, most numerical bootstrap studies have focused on four-point functions of three or fewer primary operators. An important problem is to understand how to incorporate more crossing constraints into bootstrap studies, with the goal of revealing the full implications of crossing symmetry and unitarity. This presents two main algorithmic challenges:
\begin{itemize}
\item Find faster optimization methods, or ways to scale up current tools.
\item Find efficient methods for exploring high-dimensional spaces of CFT data.
\end{itemize}
Below, we describe previous approaches to these challenges and highlight opportunities for future progress.

\subsection{Convex Optimization Methods}

One of the pioneering observations of \cite{Rattazzi:2008pe} is that the constraints of crossing symmetry and unitarity can be explored with convex optimization. By applying linear functionals to the crossing equations and using unitarity, one obtains an infinite set of linear inequalities on an infinite number of variables. To derive bounds on CFT data, one  searches for functionals that optimize various quantities, subject to the given inequalities.\footnote{This approach closely mirrors an earlier approach to bounds on sphere packings due to Cohn and Elkies \cite{cohnelkies}.} The initial work \cite{Rattazzi:2008pe} performed optimizations using an off-the-shelf linear programming solver. Subsequently, in~\cite{Poland:2011ey,Kos:2013tga,Kos:2014bka} it was demonstrated that general bootstrap problems can be phrased as semidefinite programs, which can be solved with interior point methods.  Interior-point algorithms for bootstrap problems must often be run with high-precision arithmetic. Although this incurs a large performance penalty relative to using machine precision, high-precision semidefinite programming methods can be more efficient and robust in practice than their linear-programming predecessors, and they have become popular for bootstrap problems.

One of the most widely-used optimizers for bootstrap computations is the semidefinite program solver \texttt{SDPB} \cite{Simmons-Duffin:2015qma}, which was specially designed for the conformal bootstrap. Recently, \texttt{SDPB} was enhanced to use the Elemental library~\cite{poulson2013elemental,petschow2013high} for distributed high precision linear algebra,  improving speed and enabling parallelization across several HPC nodes. This new version of \texttt{SDPB}, described in \cite{Landry:2019qug} and available at~\cite{sdpb}, greatly expands the reach and applicability of numerical bootstrap techniques. Nevertheless, there are many opportunities for further scaling and improving this code, including implementing more sophisticated distributed linear algebra routines, improving memory management to allow scaling past $\sim 500$ cores, finding ways to reduce precision requirements, and leveraging new hardware like GPUs and FPGAs. These engineering challenges are tightly coupled to physics: an increase in the scale of solvable semidefinite programs could allow one to explore larger systems of crossing equations and thereby access new CFTs.

Bootstrap researchers have also explored other convex optimization techniques. For example, the popular package \texttt{JuliBootS}~\cite{Paulos:2014vya} implements a modified simplex algorithm (based on \cite{El-Showk:2014dwa}) for solving semi-infinite programs. The program \texttt{outer\_limits} (included with {\tt SDPB}) is an experimental implementation of outer approximation methods for bootstrap computations. A virtue of these methods is that they can easily incorporate more sophisticated spaces of functionals, see e.g.\ \cite{Paulos:2019fkw} in 1-dimensional CFTs. For higher-dimensional CFTs, it would be interesting to adapt these methods to incorporate functionals that are known to carry interesting analytical information, such as dispersive functionals \cite{Mazac:2019shk,Penedones:2019tng,Kologlu:2019bco,Carmi:2019cub,Caron-Huot:2020adz}. It will also be important to explore whether new convex optimization techniques can reduce precision requirements in bootstrap computations.

\subsection{Nonlinear Optimization Methods}

In some cases, detailed knowledge of the structure of a bootstrap problem can lead to much more efficient solution methods. For example, in \cite{Afkhami-Jeddi:2019zci} the spinless modular bootstrap \cite{Hellerman:2009bu} was re-cast as a nonlinear optimization problem that can be solved efficiently with Newton's method. A key ingredient in \cite{Afkhami-Jeddi:2019zci} was an accurate ansatz for the {\it extremal spectrum} (the set of scaling dimensions that appear in an optimal solution), which provided a good starting point for Newton's method. The resulting approach is orders of magnitude faster than {\tt SDPB} applied to the same problem \cite{Collier:2016cls}. 

To apply these methods more generally, an important challenge is to generate similarly clever ansatze for extremal spectra of higher-dimensional CFTs. This may require a more systematic understanding of the boundary of the space of unitary CFT data \cite{Arkani-Hamed:2018ign}. Extremal spectra of higher-dimensional CFTs have been explored using the so-called {\it extremal functional method\/} in a number of works \cite{Poland:2010wg,El-Showk:2012vjm,El-Showk:2014dwa,El-Showk:2016mxr,Simmons-Duffin:2016wlq,Liu:2020tpf,Caron-Huot:2020ouj}, revealing intriguing but currently poorly-understood phenomena in the extremal spectrum near known CFTs. A clearer understanding of these phenomena may be important. Note that we find it unlikely that a nontrivial interacting CFT could be solved exactly by finding an ansatz for an extremal functional. For example, the spectrum of such a CFT likely exhibits elements of chaos at high scaling dimensions, and it is difficult to imagine an extremal functional with chaotic zeros. Nevertheless, perhaps this approach could be used to solve bootstrap problems for a finite number of four-point correlators to high precision.

Interpreting the crossing equations as nonlinear constraints on scaling dimensions has inspired other numerical approaches. For example, \cite{Gliozzi:2013ysa} showed how approximate CFT spectra can sometimes be obtained from truncated systems of crossing equations. Other works have explored how solutions to optimization problems can be continuously deformed to generate new solutions to crossing \cite{El-Showk:2016mxr,Afkhami-Jeddi:2021iuw}. Importantly, by not relying on positivity, these methods can be applied to non-unitary theories, or situations where the crossing equations do not enjoy manifest positivity \cite{Liendo:2012hy,Gliozzi:2014jsa,Gliozzi:2015qsa,Iliesiu:2018fao,Iliesiu:2018zlz}.

\subsection{Searching the Space of CFT Data}

A major challenge for the modern numerical bootstrap is the need to search through high-dimensional spaces of scaling dimensions and OPE coefficients (CFT data). In favorable situations, scaling dimensions and OPE coefficients of operators in the correlators being studied --- known as ``external" CFT data --- lie in allowed ``islands." To make physical predictions, we must identify the boundaries of these islands.\footnote{It can happen that higher-dimensional islands are ``smaller" along all their axes than lower-dimensional islands \cite{Kos:2016ysd} --- hence it can be physically important to compute islands in higher dimensions.} In other words, it is not enough to solve an individual convex optimization problem for some given external CFT data --- we must scan over the space of external CFT data, in principle solving an optimization problem at each point, to map out theory space. This challenge becomes more acute as one studies larger systems of correlation functions. The full problem of finding boundaries of allowed ``islands" is intrinsically nonlinear and has no known convex relaxation. 

 A compound approach to this problem was introduced in~\cite{Liu:2020tpf} (and subsequently used in~\cite{Chester:2020iyt}), which consists of a ``Delaunay triangulation search" in external dimension space and the use of a ``cutting surface" algorithm in external OPE coefficient space. Both types of searches can be made more efficient with ``hot-starting", where one uses the final state of the solver from a previous calculation as the initial state (i.e.~initial \texttt{SDPB} checkpoint file) in a new calculation \cite{Go:2019lke}. 

The idea of the cutting surface algorithm is to take advantage of the fact that each computed functional, designed to exclude some chosen values of scaling dimensions and OPE coefficients, actually excludes a wide region of OPE coefficient space (without significant additional computation). This region is defined by a quadratic inequality. By testing successive points in OPE coefficient space, and taking intersections of multiple such regions, one can efficiently narrow down possible locations of allowed points. The works~\cite{Liu:2020tpf,Chester:2020iyt} carried out this strategy in a 3-dimensional space of OPE coefficients, and in practice found that $\sim 30$ or so tests were sufficient to rule out all of OPE coefficient space or find an allowed point.

The basic idea of the Delaunay triangulation search is to start with a list of scaling dimensions $\{\Delta_i\}$ that are known to be either allowed or disallowed. Then one can perform a Delaunay triangulation of these points, focus on triangles that contain both allowed and disallowed points, and select suitable points inside those triangles to test. By iterating, one can find the boundary of the allowed region in dimension space. This is essentially a higher-dimensional version of binary search.

A weakness of the Delaunay triangulation search is that it treats the convex optimization solver as an oracle giving boolean answers (``allowed" or ``disallowed"). In general, the solution to a convex optimization problem carries much more than a single bit of data, and it is thus desirable to extract more information about the space of allowed CFT data from each optimization. This is the motivation behind the introduction of ``navigator functions" in~\cite{Reehorst:2021ykw}. The basic idea is to define a continuous and locally differentiable function $\mathcal{N}(x)$ of the CFT data $x$ that is negative in the allowed region and positive in the disallowed region. By minimizing the navigator function (e.g. with gradient descent or the more sophisticated BFGS algorithm) one can efficiently locate an allowed region. A modification of this idea can also find the extreme value of some scaling dimension over the allowed region, giving an attractive approach to determining the ranges of allowed CFT data over complicated multidimensional spaces. Navigator function methods hold promise for searching CFT data in much higher dimensional spaces than have been previously attempted \cite{Reehorst:2021hmp}. It will be important to explore navigator functions for different CFTs, finding ways to deal with local minima, discontinuities in derivatives, and other features that may hinder efficient optimization.

\subsection{Hybrid Numerical/Analytical Approaches}

The emergence of the numerical bootstrap also initiated a renaissance in analytical bootstrap methods, see the Snowmass white paper on the analytical conformal bootstrap \cite{Hartman:2022zik}. Among other results, the analytic bootstrap has established the existence of {\it multi-twist\/} operators, which are families of operators whose scaling dimensions vary in a predictable way at large spin \cite{Fitzpatrick:2012yx,Komargodski:2012ek}. Multi-twist operators can be seen in numerical bootstrap computations. For example, of the hundreds of operators in the 3d Ising and $O(2)$ models revealed using extremal functional methods, most are double-twist \cite{Simmons-Duffin:2016wlq,Liu:2020tpf}. This suggests that by directly inputting analytical information about double-twist operators (and not forcing the optimization solver to reproduce known results), one may be able to extend the reach of numerical methods; see \cite{Su:2022xnj} for promising progress in this direction.

\section{Computational Resources}

Many simple conformal bootstrap calculations can be performed on a laptop, if one restricts to low-dimensional search spaces. On the other hand, the \texttt{SDPB} software has been designed to efficiently take advantage of parallelization on HPC clusters. E.g., on the XSEDE Comet cluster tests have shown that it has good scaling behavior up to 32 nodes (24 cores each). Many medium-to-large scale computations have now been performed using this software on various university-based and national clusters. One example is the computation of scaling dimensions in the 3d Ising model~\cite{Kos:2016ysd}, which used $\sim 200k$ CPU-hours spread over a period of 3 weeks. The most computationally-intensive runs (e.g. for the O(3) model~\cite{Chester:2020iyt}) have used a few million CPU-hours, which go into testing order $O(100-1000)$ points in some space of CFT data, many of which can be computed in parallel, to map out an allowed region. Modern software tools (e.g.~\texttt{hyperion-bootstrap} or~\texttt{simpleboot}, see appendix~\ref{app:software}) have been designed to efficiently manage these searches on HPC clusters, utilizing the various search strategies (Delaunay search, cutting surface algorithm, navigator method, etc) described above.

The computation time scales with a number of factors, including the number of crossing equations, the size of the search space (typically parametrized using a derivative order $\Lambda$), the needed precision, the order of the conformal block approximation, and the number of constraints imposed (e.g., how many spins are included). There are often interesting tradeoffs to be made in deciding whether to add new crossing equations or push to higher values of $\Lambda$. The best strategy is often closely tied to interesting theoretical questions about which OPE channels are accessed and whether imposing gaps in those channels are effective in isolating physical CFTs.

One current computational bottleneck is the need for a significant amount of memory. To give a concrete example, we are currently running numerical bootstrap computations (targeting the Gross-Neveu-Yukawa models~\cite{GNYtoappear}) on the XSEDE Expanse cluster, where standard compute nodes consist of two 64-core AMD EPYC 7742 processors with 256 GB of DDR4 memory. However, even computations at moderate derivative order $\Lambda = 27, 35$ run into memory constraints and can require more than the available 2 GB/core, leading to significant inefficiencies in our ability to take advantage of the cluster. The need for a large amount of memory/core is closely tied to the need for computations to run at a high precision ($\simeq 1000$ binary digits), reflecting large cancellations which occur when computing functionals with the desired positivity properties. In the future it will be important to identify ways to reduce the needed precision and memory/core in large-scale bootstrap computations.

\section{Outlook}

Over the last $\sim 14$ years the numerical conformal bootstrap has
grown from a clever idea into a powerful set of machinery for learning about strongly-coupled conformal field theories. On the one hand, a
relatively mature toolchain has been developed which allows for
efficient explorations of the rigorous bounds that can be obtained by
applying semidefinite programming methods to sets of CFT crossing
equations. We are still only beginning to mine the quantitative
constraints and qualitative insights that can be obtained using these
methods, particularly when combined with insights from perturbative and lattice approaches. On the other hand, powerful new ideas keep emerging, most
recently the navigator methods and numerical/analytical hybrid
approaches, which help to illustrate that the numerical bootstrap may
still see orders of magnitude improvements in the near future.

Many open questions remain for the numerical bootstrap: What is the
most efficient basis of functionals? How does convergence of the
island sizes relate to the chosen crossing relations and gaps imposed? How can one eliminate the need for high precision computations? What is the best
way to incorporate analytical insights while preserving rigorous error
bars? What is the optimal approach to mapping out high-dimension parameter spaces and placing constraints on subleading operator dimensions?

We are optimistic that making further progress on these questions will
enable a new round of fundamental discoveries in theoretical physics, including the
numerical classification of CFTs with a small number of relevant
operators, definitive answers to longstanding questions about
conformal windows of gauge theories in 3 and 4 dimensions, more-or-less complete numerical solutions of the known maximally-supersymmetric CFTs (and their holographic duals), and new robust starting points for Hamiltonian truncation studies of gapped phases~\cite{Fitzpatrick:2022dwq}. Importantly, the numerical bootstrap can also be used as a discovery tool for finding previously unknown CFTs and as well as for identifying possible analytical solutions of known ones.

Finally, it is worth highlighting the many other bootstraps across both physics and mathematics where substantive progress has been recently made using numerical methods, including the S-matrix bootstrap~\cite{Kruczenski:2022lot}, the matrix model bootstrap~\cite{Lin:2020mme,Han:2020bkb,Kazakov:2021lel}, the spectral geometry bootstrap~\cite{Bonifacio:2020xoc,Kravchuk:2021akc,Bonifacio:2021aqf}, and the 2d modular bootstrap~\cite{Hellerman:2009bu} along with its connection to sphere packing~\cite{Hartman:2019pcd, Afkhami-Jeddi:2020hde}. Given the similarity of these problems and the methods used to tackle them, insights in one realm can have an immediate impact on the others. It is an exciting time to be bootstrapping!

\section*{Acknowledgements}
We thank Nima Afkhami-Jeddi, Shai Chester, Rajeev Erramilli, Tom Hartman, Yin-Chen He, Luca Iliesiu, Petr Kravchuk, Walter Landry, Zhijin Li, Aike Liu, Junyu Liu, Matthew Mitchell, Slava Rychkov, Ning Su, Amir Tajdini, Leonardo Rastelli, Balt van Rees, and Yuan Xin for discussions. DP is supported by Simons Foundation grant 488651 (Simons Collaboration on the Nonperturbative Bootstrap) and DOE grant no.\ DE-SC0017660. DSD is supported by Simons Foundation grant 488657 (Simons Collaboration on the Nonperturbative Bootstrap) and a DOE Early Career Award under grant no.\ DE-SC0019085.

\appendix 

\section{Software for Setting Up Bootstrap Computations}
\label{app:software}

Several different libraries have been developed to help set up bootstrap problems that can be solved by \texttt{SDPB}~\cite{sdpb} or other methods. Publicly available codes include:

\begin{itemize}
\item \texttt{JuliBootS}~\cite{Paulos:2014vya}: A Julia-based package for solving bootstrap problems using linear programming methods
\item \texttt{PyCFTBoot}~\cite{Behan:2016dtz}: A Python-based interface for setting up bootstrap computations for use with \texttt{SDPB}
\item \texttt{cboot}~\cite{cboot}: A Sage module for setting up bootstrap computations for use with \texttt{SDPB}
\item \texttt{simpleboot}~\cite{simpleboot}: A \texttt{Mathematica} framework for setting up bootstrap computations for use with \texttt{SDPB}
\item \texttt{hyperion-bootstrap}~\cite{hyperion-bootstrap}: A \texttt{Haskell} library for setting up numerical conformal bootstrap computations for use with \texttt{SDPB}
\end{itemize}

\noindent In addition various other ``helper" software tools have been developed, including:

\begin{itemize}
\item \texttt{scalar\_blocks}~\cite{scalarblocks}: Computes conformal blocks appearing in 4-point functions of scalar operators in any dimension
\item \texttt{blocks\_3d}~\cite{Erramilli:2020rlr}: Computes conformal blocks appearing in 4-point functions of spinning operators in 3d
\item \texttt{CFTs4D}~\cite{Cuomo:2017wme}: \texttt{Mathematica} package for conformal bootstrap calculations in 4d
\item \texttt{virasoro}~\cite{virasoro}: A program for numerical computation of Virasoro block coefficients in 2d CFT
\item \texttt{autoboot}~\cite{Go:2019lke}: A \texttt{Mathematica} package for generating mixed-correlator bootstrap equations given a global symmetry group and operator representations
\item \texttt{quadratic-net}~\cite{Liu:2020tpf, quadraticnet}: A \texttt{Haskell} library and standalone executable for solving quadratically constrained problems, used in cutting surface algorithm
\item \texttt{tiptop}~\cite{Chester:2020iyt, tiptop}: Generates successive points for searching for the maximum of an N-dimensional function in N+1-dimensional space
\item \texttt{spectrum}: Extracts the extremal spectrum from an SDP solution (included in \texttt{SDPB}) following an algorithm introduced in~\cite{Komargodski:2016auf,Simmons-Duffin:2016wlq}
\item \texttt{approx\_objective}: Derivation of linear and quadratic variations of the objective function, used in Navigator approach~\cite{Reehorst:2021ykw} (included in \texttt{SDPB})
\end{itemize}

\bibliographystyle{JHEP}
\bibliography{refs}

\providecommand{\href}[2]{#2}\begingroup\raggedright\begin{thebibliography}{100}

\bibitem{Belavin:1984vu}
A.~A. Belavin, A.~M. Polyakov and A.~B. Zamolodchikov, \emph{{Infinite
  conformal symmetry in two-dimensional quantum field theory}},
  \href{http://dx.doi.org/10.1016/0550-3213(84)90052-X}{\emph{Nucl. Phys.} {\bf
  B241} (1984) 333--380}.

\bibitem{Rattazzi:2008pe}
R.~Rattazzi, V.~S. Rychkov, E.~Tonni and A.~Vichi, \emph{{Bounding scalar
  operator dimensions in 4D CFT}},
  \href{http://dx.doi.org/10.1088/1126-6708/2008/12/031}{\emph{JHEP} {\bf 12}
  (2008) 031}, [\href{https://arxiv.org/abs/0807.0004}{{\tt 0807.0004}}].

\bibitem{Poland:2018epd}
D.~Poland, S.~Rychkov and A.~Vichi, \emph{{The Conformal Bootstrap: Theory,
  Numerical Techniques, and Applications}},
  \href{http://dx.doi.org/10.1103/RevModPhys.91.015002}{\emph{Rev. Mod. Phys.}
  {\bf 91} (2019) 15002}, [\href{https://arxiv.org/abs/1805.04405}{{\tt
  1805.04405}}].

\bibitem{Hartman:2022zik}
T.~Hartman, D.~Mazac, D.~Simmons-Duffin and A.~Zhiboedov, \emph{{Snowmass White
  Paper: The Analytic Conformal Bootstrap}},  in \emph{{2022 Snowmass Summer
  Study}}, 2, 2022.
\newblock \href{https://arxiv.org/abs/2202.11012}{{\tt 2202.11012}}.

\bibitem{Gopakumar:2022kof}
R.~Gopakumar, E.~Perlmutter, S.~S. Pufu and X.~Yin, \emph{{Snowmass White
  Paper: Bootstrapping String Theory}},
  \href{https://arxiv.org/abs/2202.07163}{{\tt 2202.07163}}.

\bibitem{Kruczenski:2022lot}
M.~Kruczenski, J.~Penedones and B.~C. van Rees, \emph{{Snowmass White Paper:
  S-matrix Bootstrap}},  \href{https://arxiv.org/abs/2203.02421}{{\tt
  2203.02421}}.

\bibitem{Argyres:2022mnu}
P.~C. Argyres, J.~J. Heckman, K.~Intriligator and M.~Martone, \emph{{Snowmass
  White Paper on SCFTs}},  \href{https://arxiv.org/abs/2202.07683}{{\tt
  2202.07683}}.

\bibitem{Fitzpatrick:2022dwq}
A.~L. Fitzpatrick and E.~Katz, \emph{{Snowmass White Paper: Hamiltonian
  Truncation}},  \href{https://arxiv.org/abs/2201.11696}{{\tt 2201.11696}}.

\bibitem{Henriksson:2022rnm}
J.~Henriksson, \emph{{The critical $O(N)$ CFT: Methods and conformal data}},
  \href{https://arxiv.org/abs/2201.09520}{{\tt 2201.09520}}.

\bibitem{El-Showk:2012cjh}
S.~El-Showk, M.~F. Paulos, D.~Poland, S.~Rychkov, D.~Simmons-Duffin and
  A.~Vichi, \emph{{Solving the 3D Ising Model with the Conformal Bootstrap}},
  \href{http://dx.doi.org/10.1103/PhysRevD.86.025022}{\emph{Phys. Rev. D} {\bf
  86} (2012) 025022}, [\href{https://arxiv.org/abs/1203.6064}{{\tt
  1203.6064}}].

\bibitem{El-Showk:2013nia}
S.~El-Showk, M.~Paulos, D.~Poland, S.~Rychkov, D.~Simmons-Duffin and A.~Vichi,
  \emph{Conformal field theories in fractional dimensions},
  \href{http://dx.doi.org/10.1103/PhysRevLett.112.141601}{\emph{Phys. Rev.
  Lett.} {\bf 112} (Apr, 2014) 141601},
  [\href{https://arxiv.org/abs/1309.5089}{{\tt 1309.5089}}].

\bibitem{El-Showk:2014dwa}
S.~El-Showk, M.~F. Paulos, D.~Poland, S.~Rychkov, D.~Simmons-Duffin and
  A.~Vichi, \emph{{Solving the 3d Ising Model with the Conformal Bootstrap II.
  c-Minimization and Precise Critical Exponents}},
  \href{http://dx.doi.org/10.1007/s10955-014-1042-7}{\emph{J. Stat. Phys.} {\bf
  157} (2014) 869}, [\href{https://arxiv.org/abs/1403.4545}{{\tt 1403.4545}}].

\bibitem{Kos:2014bka}
F.~Kos, D.~Poland and D.~Simmons-Duffin, \emph{{Bootstrapping Mixed Correlators
  in the 3D Ising Model}},
  \href{http://dx.doi.org/10.1007/JHEP11(2014)109}{\emph{JHEP} {\bf 11} (2014)
  109}, [\href{https://arxiv.org/abs/1406.4858}{{\tt 1406.4858}}].

\bibitem{Kos:2016ysd}
F.~Kos, D.~Poland, D.~Simmons-Duffin and A.~Vichi, \emph{{Precision islands in
  the Ising and O(N ) models}},
  \href{http://dx.doi.org/10.1007/JHEP08(2016)036}{\emph{JHEP} {\bf 08} (2016)
  036}, [\href{https://arxiv.org/abs/1603.04436}{{\tt 1603.04436}}].

\bibitem{Behan:2017dwr}
C.~Behan, L.~Rastelli, S.~Rychkov and B.~Zan, \emph{{Long-range critical
  exponents near the short-range crossover}},
  \href{http://dx.doi.org/10.1103/PhysRevLett.118.241601}{\emph{Phys. Rev.
  Lett.} {\bf 118} (2017) 241601},
  [\href{https://arxiv.org/abs/1703.03430}{{\tt 1703.03430}}].

\bibitem{Behan:2017emf}
C.~Behan, L.~Rastelli, S.~Rychkov and B.~Zan, \emph{{A scaling theory for the
  long-range to short-range crossover and an infrared duality}},
  \href{http://dx.doi.org/10.1088/1751-8121/aa8099}{\emph{J. Phys.} {\bf A50}
  (2017) 354002}, [\href{https://arxiv.org/abs/1703.05325}{{\tt 1703.05325}}].

\bibitem{Behan:2018hfx}
C.~Behan, \emph{{Bootstrapping the long-range Ising model in three
  dimensions}}, \href{http://dx.doi.org/10.1088/1751-8121/aafd1b}{\emph{J.
  Phys.} {\bf A52} (2019) 075401},
  [\href{https://arxiv.org/abs/1810.07199}{{\tt 1810.07199}}].

\bibitem{Simmons-Duffin:2016wlq}
D.~Simmons-Duffin, \emph{{The Lightcone Bootstrap and the Spectrum of the 3d
  Ising CFT}}, \href{http://dx.doi.org/10.1007/JHEP03(2017)086}{\emph{JHEP}
  {\bf 03} (2017) 086}, [\href{https://arxiv.org/abs/1612.08471}{{\tt
  1612.08471}}].

\bibitem{Albayrak:2019gnz}
S.~Albayrak, D.~Meltzer and D.~Poland, \emph{{More Analytic Bootstrap:
  Nonperturbative Effects and Fermions}},
  \href{http://dx.doi.org/10.1007/JHEP08(2019)040}{\emph{JHEP} {\bf 08} (2019)
  040}, [\href{https://arxiv.org/abs/1904.00032}{{\tt 1904.00032}}].

\bibitem{Caron-Huot:2020ouj}
S.~Caron-Huot, Y.~Gobeil and Z.~Zahraee, \emph{{The leading trajectory in the
  2+1D Ising CFT}},  \href{https://arxiv.org/abs/2007.11647}{{\tt 2007.11647}}.

\bibitem{Reehorst:2021hmp}
M.~Reehorst, \emph{{Rigorous bounds on irrelevant operators in the 3d Ising
  model CFT}},  \href{https://arxiv.org/abs/2111.12093}{{\tt 2111.12093}}.

\bibitem{Su:2022xnj}
N.~Su, \emph{{The Hybrid Bootstrap}},
  \href{https://arxiv.org/abs/2202.07607}{{\tt 2202.07607}}.

\bibitem{PhysRevB.44.6883}
M.-C. Cha, M.~P.~A. Fisher, S.~M. Girvin, M.~Wallin and A.~P. Young,
  \emph{Universal conductivity of two-dimensional films at the
  superconductor-insulator transition},
  \href{http://dx.doi.org/10.1103/PhysRevB.44.6883}{\emph{Phys. Rev. B} {\bf
  44} (Oct, 1991) 6883--6902}.

\bibitem{PhysRevLett.95.180603}
J.~\ifmmode~\check{S}\else \v{S}\fi{}makov and E.~S{\o}rensen, \emph{Universal
  scaling of the conductivity at the superfluid-insulator phase transition},
  \href{http://dx.doi.org/10.1103/PhysRevLett.95.180603}{\emph{Phys. Rev.
  Lett.} {\bf 95} (Oct, 2005) 180603}.

\bibitem{PhysRevB.68.174518}
J.~A. Lipa, J.~A. Nissen, D.~A. Stricker, D.~R. Swanson and T.~C.~P. Chui,
  \emph{Specific heat of liquid helium in zero gravity very near the lambda
  point}, \href{http://dx.doi.org/10.1103/PhysRevB.68.174518}{\emph{Phys. Rev.
  B} {\bf 68} (Nov, 2003) 174518}.

\bibitem{PhysRevB.74.144506}
M.~Campostrini, M.~Hasenbusch, A.~Pelissetto and E.~Vicari, \emph{Theoretical
  estimates of the critical exponents of the superfluid transition in
  $^{4}\mathrm{He}$ by lattice methods},
  \href{http://dx.doi.org/10.1103/PhysRevB.74.144506}{\emph{Phys. Rev. B} {\bf
  74} (Oct, 2006) 144506}.

\bibitem{Go:2019lke}
M.~Go and Y.~Tachikawa, \emph{{autoboot: A generator of bootstrap equations
  with global symmetry}},
  \href{http://dx.doi.org/10.1007/JHEP06(2019)084}{\emph{JHEP} {\bf 06} (2019)
  084}, [\href{https://arxiv.org/abs/1903.10522}{{\tt 1903.10522}}].

\bibitem{Chester:2019ifh}
S.~M. Chester, W.~Landry, J.~Liu, D.~Poland, D.~Simmons-Duffin, N.~Su et~al.,
  \emph{{Carving out OPE space and precise $O(2)$ model critical exponents}},
  \href{http://dx.doi.org/10.1007/JHEP06(2020)142}{\emph{JHEP} {\bf 06} (2020)
  142}, [\href{https://arxiv.org/abs/1912.03324}{{\tt 1912.03324}}].

\bibitem{doi:10.1080/00150198908245184}
I.~O. Mayer, A.~I. Sokolov and B.~N. Shalayev, \emph{Critical exponents for
  cubic and impure uniaxial crystals: most accurate (?) theoretical values},
  \href{http://dx.doi.org/10.1080/00150198908245184}{\emph{Ferroelectrics} {\bf
  95} (1989) 93--96}.

\bibitem{PhysRevB.61.14660}
K.~B. Varnashev, \emph{Stability of a cubic fixed point in three dimensions:
  Critical exponents for generic n},
  \href{http://dx.doi.org/10.1103/PhysRevB.61.14660}{\emph{Phys. Rev. B} {\bf
  61} (Jun, 2000) 14660--14674}.

\bibitem{PhysRevB.61.15130}
D.~V. Pakhnin and A.~I. Sokolov, \emph{Five-loop renormalization-group
  expansions for the three-dimensional n-vector cubic model and critical
  exponents for impure ising systems},
  \href{http://dx.doi.org/10.1103/PhysRevB.61.15130}{\emph{Phys. Rev. B} {\bf
  61} (Jun, 2000) 15130--15135}.

\bibitem{Caselle_1998}
M.~Caselle and M.~Hasenbusch, \emph{The stability of the o(n) invariant fixed
  point in three dimensions},
  \href{http://dx.doi.org/10.1088/0305-4470/31/20/004}{\emph{Journal of Physics
  A: Mathematical and General} {\bf 31} (may, 1998) 4603--4617}.

\bibitem{PhysRevE.94.042107}
A.~Kudlis and A.~I. Sokolov, \emph{Anisotropy of a cubic ferromagnet at
  criticality}, \href{http://dx.doi.org/10.1103/PhysRevE.94.042107}{\emph{Phys.
  Rev. E} {\bf 94} (Oct, 2016) 042107}.

\bibitem{PhysRevB.62.12195}
R.~Folk, Y.~Holovatch and T.~Yavors'kii,
  \emph{Pseudo-$\ensuremath{\varepsilon}$ expansion of six-loop
  renormalization-group functions of an anisotropic cubic model},
  \href{http://dx.doi.org/10.1103/PhysRevB.62.12195}{\emph{Phys. Rev. B} {\bf
  62} (Nov, 2000) 12195--12200}.

\bibitem{PhysRevB.27.609}
G.~Jug, \emph{Critical behavior of disordered spin systems in two and three
  dimensions}, \href{http://dx.doi.org/10.1103/PhysRevB.27.609}{\emph{Phys.
  Rev. B} {\bf 27} (Jan, 1983) 609--612}.

\bibitem{PhysRevLett.33.813}
D.~R. Nelson, J.~M. Kosterlitz and M.~E. Fisher, \emph{Renormalization-group
  analysis of bicritical and tetracritical points},
  \href{http://dx.doi.org/10.1103/PhysRevLett.33.813}{\emph{Phys. Rev. Lett.}
  {\bf 33} (Sep, 1974) 813--817}.

\bibitem{SHALAEV1997105}
B.~Shalaev, S.~Antonenko and A.~Sokolov, \emph{Five-loop $\epsilon$ for random
  ising model and marginal spin dimensionality for cubic systems},
  \href{http://dx.doi.org/https://doi.org/10.1016/S0375-9601(97)00225-9}{\emph{Physics
  Letters A} {\bf 230} (1997) 105 -- 110}.

\bibitem{KLEINERT1995284}
H.~Kleinert and V.~Schulte-Frohlinde, \emph{Exact five-loop renormalization
  group functions of $\theta^4$-theory with o(n)-symmetric and cubic
  interactions. critical exponents up to $\epsilon^5$},
  \href{http://dx.doi.org/https://doi.org/10.1016/0370-2693(94)01377-O}{\emph{Physics
  Letters B} {\bf 342} (1995) 284 -- 296}.

\bibitem{PhysRevB.56.14428}
H.~Kleinert, S.~Thoms and V.~Schulte-Frohlinde, \emph{Stability of a
  three-dimensional cubic fixed point in the two-coupling-constant
  ${\ensuremath{\varphi}}^{4}$ theory},
  \href{http://dx.doi.org/10.1103/PhysRevB.56.14428}{\emph{Phys. Rev. B} {\bf
  56} (Dec, 1997) 14428--14434}.

\bibitem{PhysRevB.65.144520}
M.~Campostrini, M.~Hasenbusch, A.~Pelissetto, P.~Rossi and E.~Vicari,
  \emph{Critical exponents and equation of state of the three-dimensional
  heisenberg universality class},
  \href{http://dx.doi.org/10.1103/PhysRevB.65.144520}{\emph{Phys. Rev. B} {\bf
  65} (Apr, 2002) 144520}.

\bibitem{PhysRevB.61.15136}
J.~Manuel~Carmona, A.~Pelissetto and E.~Vicari, \emph{$n$-component
  ginzburg-landau hamiltonian with cubic anisotropy: A six-loop study},
  \href{http://dx.doi.org/10.1103/PhysRevB.61.15136}{\emph{Phys. Rev. B} {\bf
  61} (Jun, 2000) 15136--15151}.

\bibitem{Pakhnin2002}
D.~V. Pakhnin, A.~I. Sokolov and B.~N. Shalaev, \emph{Nonlinear
  susceptibilities of a weakly disordered uniaxial ferromagnet in the critical
  region}, \href{http://dx.doi.org/10.1134/1.1490005}{\emph{Journal of
  Experimental and Theoretical Physics Letters} {\bf 75} (Apr, 2002) 387--390}.

\bibitem{Chester:2020iyt}
S.~M. Chester, W.~Landry, J.~Liu, D.~Poland, D.~Simmons-Duffin, N.~Su et~al.,
  \emph{{Bootstrapping Heisenberg magnets and their cubic instability}},
  \href{http://dx.doi.org/10.1103/PhysRevD.104.105013}{\emph{Phys. Rev. D} {\bf
  104} (2021) 105013}, [\href{https://arxiv.org/abs/2011.14647}{{\tt
  2011.14647}}].

\bibitem{Rong:2017cow}
J.~Rong and N.~Su, \emph{{Scalar CFTs and Their Large N Limits}},
  \href{http://dx.doi.org/10.1007/JHEP09(2018)103}{\emph{JHEP} {\bf 09} (2018)
  103}, [\href{https://arxiv.org/abs/1712.00985}{{\tt 1712.00985}}].

\bibitem{Stergiou:2018gjj}
A.~Stergiou, \emph{{Bootstrapping hypercubic and hypertetrahedral theories in
  three dimensions}},
  \href{http://dx.doi.org/10.1007/JHEP05(2018)035}{\emph{JHEP} {\bf 05} (2018)
  035}, [\href{https://arxiv.org/abs/1801.07127}{{\tt 1801.07127}}].

\bibitem{Kousvos:2018rhl}
S.~R. Kousvos and A.~Stergiou, \emph{{Bootstrapping Mixed Correlators in
  Three-Dimensional Cubic Theories}},
  \href{http://dx.doi.org/10.21468/SciPostPhys.6.3.035}{\emph{SciPost Phys.}
  {\bf 6} (2019) 035}, [\href{https://arxiv.org/abs/1810.10015}{{\tt
  1810.10015}}].

\bibitem{Kousvos:2019hgc}
S.~R. Kousvos and A.~Stergiou, \emph{{Bootstrapping Mixed Correlators in
  Three-Dimensional Cubic Theories II}},
  \href{http://dx.doi.org/10.21468/SciPostPhys.8.6.085}{\emph{SciPost Phys.}
  {\bf 8} (2020) 085}, [\href{https://arxiv.org/abs/1911.00522}{{\tt
  1911.00522}}].

\bibitem{Nakayama:2014lva}
Y.~Nakayama and T.~Ohtsuki, \emph{{Approaching the conformal window of
  $O(n)\times O(m)$ symmetric Landau-Ginzburg models using the conformal
  bootstrap}}, \href{http://dx.doi.org/10.1103/PhysRevD.89.126009}{\emph{Phys.
  Rev.} {\bf D89} (2014) 126009}, [\href{https://arxiv.org/abs/1404.0489}{{\tt
  1404.0489}}].

\bibitem{Nakayama:2014sba}
Y.~Nakayama and T.~Ohtsuki, \emph{{Bootstrapping phase transitions in QCD and
  frustrated spin systems}},
  \href{http://dx.doi.org/10.1103/PhysRevD.91.021901}{\emph{Phys. Rev.} {\bf
  D91} (2015) 021901}, [\href{https://arxiv.org/abs/1407.6195}{{\tt
  1407.6195}}].

\bibitem{Henriksson:2020fqi}
J.~Henriksson, S.~R. Kousvos and A.~Stergiou, \emph{{Analytic and Numerical
  Bootstrap of CFTs with $O(m)\times O(n)$ Global Symmetry in 3D}},
  \href{http://dx.doi.org/10.21468/SciPostPhys.9.3.035}{\emph{SciPost Phys.}
  {\bf 9} (2020) 035}, [\href{https://arxiv.org/abs/2004.14388}{{\tt
  2004.14388}}].

\bibitem{Reehorst:2020phk}
M.~Reehorst, M.~Refinetti and A.~Vichi, \emph{{Bootstrapping traceless
  symmetric $O(N)$ scalars}},  \href{https://arxiv.org/abs/2012.08533}{{\tt
  2012.08533}}.

\bibitem{Stergiou:2019dcv}
A.~Stergiou, \emph{{Bootstrapping MN and Tetragonal CFTs in Three Dimensions}},
  \href{http://dx.doi.org/10.21468/SciPostPhys.7.1.010}{\emph{SciPost Phys.}
  {\bf 7} (2019) 010}, [\href{https://arxiv.org/abs/1904.00017}{{\tt
  1904.00017}}].

\bibitem{Henriksson:2021lwn}
J.~Henriksson and A.~Stergiou, \emph{{Perturbative and Nonperturbative Studies
  of CFTs with MN Global Symmetry}},
  \href{http://dx.doi.org/10.21468/SciPostPhys.11.1.015}{\emph{SciPost Phys.}
  {\bf 11} (2021) 015}, [\href{https://arxiv.org/abs/2101.08788}{{\tt
  2101.08788}}].

\bibitem{Kousvos:2021rar}
S.~R. Kousvos and A.~Stergiou, \emph{{Bootstrapping Mixed MN Correlators in
  3D}},  \href{https://arxiv.org/abs/2112.03919}{{\tt 2112.03919}}.

\bibitem{Dymarsky:2017xzb}
A.~Dymarsky, J.~Penedones, E.~Trevisani and A.~Vichi, \emph{{Charting the space
  of 3D CFTs with a continuous global symmetry}},
  \href{http://dx.doi.org/10.1007/JHEP05(2019)098}{\emph{JHEP} {\bf 05} (2019)
  098}, [\href{https://arxiv.org/abs/1705.04278}{{\tt 1705.04278}}].

\bibitem{Reehorst:2019pzi}
M.~Reehorst, E.~Trevisani and A.~Vichi, \emph{{Mixed Scalar-Current bootstrap
  in three dimensions}},
  \href{http://dx.doi.org/10.1007/JHEP12(2020)156}{\emph{JHEP} {\bf 12} (2020)
  156}, [\href{https://arxiv.org/abs/1911.05747}{{\tt 1911.05747}}].

\bibitem{Dymarsky:2017yzx}
A.~Dymarsky, F.~Kos, P.~Kravchuk, D.~Poland and D.~Simmons-Duffin, \emph{{The
  3d Stress-Tensor Bootstrap}},
  \href{http://dx.doi.org/10.1007/JHEP02(2018)164}{\emph{JHEP} {\bf 02} (2018)
  164}, [\href{https://arxiv.org/abs/1708.05718}{{\tt 1708.05718}}].

\bibitem{Hofman:2008ar}
D.~M. Hofman and J.~Maldacena, \emph{{Conformal collider physics: Energy and
  charge correlations}},
  \href{http://dx.doi.org/10.1088/1126-6708/2008/05/012}{\emph{JHEP} {\bf 05}
  (2008) 012}, [\href{https://arxiv.org/abs/0803.1467}{{\tt 0803.1467}}].

\bibitem{Kologlu:2019mfz}
M.~Kologlu, P.~Kravchuk, D.~Simmons-Duffin and A.~Zhiboedov, \emph{{The
  light-ray OPE and conformal colliders}},
  \href{http://dx.doi.org/10.1007/JHEP01(2021)128}{\emph{JHEP} {\bf 01} (2021)
  128}, [\href{https://arxiv.org/abs/1905.01311}{{\tt 1905.01311}}].

\bibitem{Dixon:2019uzg}
L.~J. Dixon, I.~Moult and H.~X. Zhu, \emph{{Collinear limit of the
  energy-energy correlator}},
  \href{http://dx.doi.org/10.1103/PhysRevD.100.014009}{\emph{Phys. Rev. D} {\bf
  100} (2019) 014009}, [\href{https://arxiv.org/abs/1905.01310}{{\tt
  1905.01310}}].

\bibitem{Chang:2020qpj}
C.-H. Chang, M.~Kologlu, P.~Kravchuk, D.~Simmons-Duffin and A.~Zhiboedov,
  \emph{{Transverse spin in the light-ray OPE}},
  \href{https://arxiv.org/abs/2010.04726}{{\tt 2010.04726}}.

\bibitem{herbut2006interactions}
I.~F. Herbut, \emph{{Interactions and phase transitions on graphene's honeycomb
  lattice}}, \href{http://dx.doi.org/10.1103/PhysRevLett.97.146401}{\emph{Phys.
  Rev. Lett.} {\bf 97} (2006) 146401},
  [\href{https://arxiv.org/abs/cond-mat/0606195}{{\tt cond-mat/0606195}}].

\bibitem{herbut2009relativistic}
I.~F. Herbut, V.~Juricic and O.~Vafek, \emph{{Relativistic Mott criticality in
  graphene}}, \href{http://dx.doi.org/10.1103/PhysRevB.80.075432}{\emph{Phys.
  Rev.} {\bf B80} (2009) 075432}, [\href{https://arxiv.org/abs/0904.1019}{{\tt
  0904.1019}}].

\bibitem{Mihaila:2017ble}
L.~N. Mihaila, N.~Zerf, B.~Ihrig, I.~F. Herbut and M.~M. Scherer,
  \emph{{Gross-Neveu-Yukawa model at three loops and Ising critical behavior of
  Dirac systems}},
  \href{http://dx.doi.org/10.1103/PhysRevB.96.165133}{\emph{Phys. Rev. B} {\bf
  96} (2017) 165133}, [\href{https://arxiv.org/abs/1703.08801}{{\tt
  1703.08801}}].

\bibitem{vojta2000quantum}
M.~Vojta, Y.~Zhang and S.~Sachdev, \emph{{Quantum Phase Transitions in d-Wave
  Superconductors}},
  \href{http://dx.doi.org/10.1103/PhysRevLett.85.4940}{\emph{Phys. Rev. Lett.}
  {\bf 85} (2000) 4940--4943}.

\bibitem{vojta2003quantum}
M.~Vojta, \emph{Quantum phase transitions}, {\emph{Reports on Progress in
  Physics} {\bf 66} (2003) 2069}.

\bibitem{Grover:2013rc}
T.~Grover, D.~N. Sheng and A.~Vishwanath, \emph{{Emergent Space-Time
  Supersymmetry at the Boundary of a Topological Phase}},
  \href{http://dx.doi.org/10.1126/science.1248253}{\emph{Science} {\bf 344}
  (2014) 280--283}, [\href{https://arxiv.org/abs/1301.7449}{{\tt 1301.7449}}].

\bibitem{Iliesiu:2015qra}
L.~Iliesiu, F.~Kos, D.~Poland, S.~S. Pufu, D.~Simmons-Duffin and R.~Yacoby,
  \emph{{Bootstrapping 3D Fermions}},
  \href{http://dx.doi.org/10.1007/JHEP03(2016)120}{\emph{JHEP} {\bf 03} (2016)
  120}, [\href{https://arxiv.org/abs/1508.00012}{{\tt 1508.00012}}].

\bibitem{Iliesiu:2017nrv}
L.~Iliesiu, F.~Kos, D.~Poland, S.~S. Pufu and D.~Simmons-Duffin,
  \emph{{Bootstrapping 3D Fermions with Global Symmetries}},
  \href{http://dx.doi.org/10.1007/JHEP01(2018)036}{\emph{JHEP} {\bf 01} (2018)
  036}, [\href{https://arxiv.org/abs/1705.03484}{{\tt 1705.03484}}].

\bibitem{Ihrig:2018hho}
B.~Ihrig, L.~N. Mihaila and M.~M. Scherer, \emph{{Critical behavior of Dirac
  fermions from perturbative renormalization}},
  \href{http://dx.doi.org/10.1103/PhysRevB.98.125109}{\emph{Phys. Rev. B} {\bf
  98} (2018) 125109}, [\href{https://arxiv.org/abs/1806.04977}{{\tt
  1806.04977}}].

\bibitem{GNYtoappear}
R.~Erramilli, L.~Iliesiu, P.~Kravchuk, A.~Liu, D.~Poland and D.~Simmons-Duffin,
  \emph{{to appear}},  2022.

\bibitem{Rong:2018okz}
J.~Rong and N.~Su, \emph{{Bootstrapping the minimal $ \mathcal{N} $ = 1
  superconformal field theory in three dimensions}},
  \href{http://dx.doi.org/10.1007/JHEP06(2021)154}{\emph{JHEP} {\bf 06} (2021)
  154}, [\href{https://arxiv.org/abs/1807.04434}{{\tt 1807.04434}}].

\bibitem{Atanasov:2018kqw}
A.~Atanasov, A.~Hillman and D.~Poland, \emph{{Bootstrapping the Minimal 3D
  SCFT}}, \href{http://dx.doi.org/10.1007/JHEP11(2018)140}{\emph{JHEP} {\bf 11}
  (2018) 140}, [\href{https://arxiv.org/abs/1807.05702}{{\tt 1807.05702}}].

\bibitem{Atanasov:2022bpi}
A.~Atanasov, A.~Hillman, D.~Poland, J.~Rong and N.~Su, \emph{{Precision
  Bootstrap for the $\mathcal{N}=1$ Super-Ising Model}},
  \href{https://arxiv.org/abs/2201.02206}{{\tt 2201.02206}}.

\bibitem{Bobev:2015vsa}
N.~Bobev, S.~El-Showk, D.~Mazac and M.~F. Paulos, \emph{{Bootstrapping the
  Three-Dimensional Supersymmetric Ising Model}},
  \href{http://dx.doi.org/10.1103/PhysRevLett.115.051601}{\emph{Phys. Rev.
  Lett.} {\bf 115} (2015) 051601},
  [\href{https://arxiv.org/abs/1502.04124}{{\tt 1502.04124}}].

\bibitem{Bobev:2015jxa}
N.~Bobev, S.~El-Showk, D.~Mazac and M.~F. Paulos, \emph{{Bootstrapping SCFTs
  with Four Supercharges}},
  \href{http://dx.doi.org/10.1007/JHEP08(2015)142}{\emph{JHEP} {\bf 08} (2015)
  142}, [\href{https://arxiv.org/abs/1503.02081}{{\tt 1503.02081}}].

\bibitem{Baggio:2017mas}
M.~Baggio, N.~Bobev, S.~M. Chester, E.~Lauria and S.~S. Pufu, \emph{{Decoding a
  Three-Dimensional Conformal Manifold}},
  \href{http://dx.doi.org/10.1007/JHEP02(2018)062}{\emph{JHEP} {\bf 02} (2018)
  062}, [\href{https://arxiv.org/abs/1712.02698}{{\tt 1712.02698}}].

\bibitem{Rong:2019qer}
J.~Rong and N.~Su, \emph{{Bootstrapping the $ \mathcal{N} $ = 1 Wess-Zumino
  models in three dimensions}},
  \href{http://dx.doi.org/10.1007/JHEP06(2021)153}{\emph{JHEP} {\bf 06} (2021)
  153}, [\href{https://arxiv.org/abs/1910.08578}{{\tt 1910.08578}}].

\bibitem{Karateev:2019pvw}
D.~Karateev, P.~Kravchuk, M.~Serone and A.~Vichi, \emph{{Fermion Conformal
  Bootstrap in 4d}},
  \href{http://dx.doi.org/10.1007/JHEP06(2019)088}{\emph{JHEP} {\bf 06} (2019)
  088}, [\href{https://arxiv.org/abs/1902.05969}{{\tt 1902.05969}}].

\bibitem{Erramilli:2020rlr}
R.~S. Erramilli, L.~V. Iliesiu, P.~Kravchuk, W.~Landry, D.~Poland and
  D.~Simmons-Duffin, \emph{{blocks\_3d: software for general 3d conformal
  blocks}}, \href{http://dx.doi.org/10.1007/JHEP11(2021)006}{\emph{JHEP} {\bf
  11} (2021) 006}, [\href{https://arxiv.org/abs/2011.01959}{{\tt 2011.01959}}].

\bibitem{Chester:2016wrc}
S.~M. Chester and S.~S. Pufu, \emph{{Towards bootstrapping QED$_{3}$}},
  \href{http://dx.doi.org/10.1007/JHEP08(2016)019}{\emph{JHEP} {\bf 08} (2016)
  019}, [\href{https://arxiv.org/abs/1601.03476}{{\tt 1601.03476}}].

\bibitem{Chester:2017vdh}
S.~M. Chester, L.~V. Iliesiu, M.~Mezei and S.~S. Pufu, \emph{{Monopole
  Operators in $U(1)$ Chern-Simons-Matter Theories}},
  \href{http://dx.doi.org/10.1007/JHEP05(2018)157}{\emph{JHEP} {\bf 05} (2018)
  157}, [\href{https://arxiv.org/abs/1710.00654}{{\tt 1710.00654}}].

\bibitem{Li:2018lyb}
Z.~Li, \emph{{Bootstrapping conformal QED$_3$ and deconfined quantum critical
  point}},  \href{https://arxiv.org/abs/1812.09281}{{\tt 1812.09281}}.

\bibitem{Li:2020bnb}
Z.~Li and D.~Poland, \emph{{Searching for gauge theories with the conformal
  bootstrap}}, \href{http://dx.doi.org/10.1007/JHEP03(2021)172}{\emph{JHEP}
  {\bf 03} (2021) 172}, [\href{https://arxiv.org/abs/2005.01721}{{\tt
  2005.01721}}].

\bibitem{He:2020azu}
Y.-C. He, J.~Rong and N.~Su, \emph{{Non-Wilson-Fisher kinks of $O(N)$ numerical
  bootstrap: from the deconfined phase transition to a putative new family of
  CFTs}}, \href{http://dx.doi.org/10.21468/SciPostPhys.10.5.115}{\emph{SciPost
  Phys.} {\bf 10} (2021) 115}, [\href{https://arxiv.org/abs/2005.04250}{{\tt
  2005.04250}}].

\bibitem{Li:2021emd}
Z.~Li, \emph{{On conformality and self-duality of $N_f=2$ QED$_3$}},
  \href{https://arxiv.org/abs/2107.09020}{{\tt 2107.09020}}.

\bibitem{He:2021sto}
Y.-C. He, J.~Rong and N.~Su, \emph{{Conformal bootstrap bounds for the $U(1)$
  Dirac spin liquid and $N=7$ Stiefel liquid}},
  \href{https://arxiv.org/abs/2107.14637}{{\tt 2107.14637}}.

\bibitem{Albayrak:2021xtd}
S.~Albayrak, R.~S. Erramilli, Z.~Li, D.~Poland and Y.~Xin, \emph{{Bootstrapping
  $N_f$=4 conformal QED$_3$}},
  \href{http://dx.doi.org/10.1103/PhysRevD.105.085008}{\emph{Phys. Rev. D} {\bf
  105} (2022) 085008}, [\href{https://arxiv.org/abs/2112.02106}{{\tt
  2112.02106}}].

\bibitem{Nakayama:2016jhq}
Y.~Nakayama and T.~Ohtsuki, \emph{{Conformal Bootstrap Dashing Hopes of
  Emergent Symmetry}},
  \href{http://dx.doi.org/10.1103/PhysRevLett.117.131601}{\emph{Phys. Rev.
  Lett.} {\bf 117} (2016) 131601},
  [\href{https://arxiv.org/abs/1602.07295}{{\tt 1602.07295}}].

\bibitem{DSD:2016}
D.~Simmons-Duffin, \emph{{Unpublished}},  2016.

\bibitem{DP:2017}
D.~Poland, \emph{{Unpublished}},  2017.

\bibitem{Iliesiu:2018}
L.~Iliesiu, \emph{{Talk at Simons Center for Geometry and Physics:
  Bootstrapping the N\'eel-VBS quantum phase transition}},  November, 2018.

\bibitem{Manenti:2021elk}
A.~Manenti and A.~Vichi, \emph{{Exploring $SU(N)$ adjoint correlators in
  $3d$}},  \href{https://arxiv.org/abs/2101.07318}{{\tt 2101.07318}}.

\bibitem{He:2021xvg}
Y.-C. He, J.~Rong and N.~Su, \emph{{A roadmap for bootstrapping critical gauge
  theories: decoupling operators of conformal field theories in $d>2$
  dimensions}},
  \href{http://dx.doi.org/10.21468/SciPostPhys.11.6.111}{\emph{SciPost Phys.}
  {\bf 11} (2021) 111}, [\href{https://arxiv.org/abs/2101.07262}{{\tt
  2101.07262}}].

\bibitem{Agmon:2017xes}
N.~B. Agmon, S.~M. Chester and S.~S. Pufu, \emph{{Solving M-theory with the
  Conformal Bootstrap}},
  \href{http://dx.doi.org/10.1007/JHEP06(2018)159}{\emph{JHEP} {\bf 06} (2018)
  159}, [\href{https://arxiv.org/abs/1711.07343}{{\tt 1711.07343}}].

\bibitem{Agmon:2019imm}
N.~B. Agmon, S.~M. Chester and S.~S. Pufu, \emph{{The M-theory Archipelago}},
  \href{http://dx.doi.org/10.1007/JHEP02(2020)010}{\emph{JHEP} {\bf 02} (2020)
  010}, [\href{https://arxiv.org/abs/1907.13222}{{\tt 1907.13222}}].

\bibitem{Binder:2020ckj}
D.~J. Binder, S.~M. Chester, M.~Jerdee and S.~S. Pufu, \emph{{The 3d $
  \mathcal{N} $ = 6 bootstrap: from higher spins to strings to membranes}},
  \href{http://dx.doi.org/10.1007/JHEP05(2021)083}{\emph{JHEP} {\bf 05} (2021)
  083}, [\href{https://arxiv.org/abs/2011.05728}{{\tt 2011.05728}}].

\bibitem{Alday:2021ymb}
L.~F. Alday, S.~M. Chester and H.~Raj, \emph{{ABJM at strong coupling from
  M-theory, localization, and Lorentzian inversion}},
  \href{http://dx.doi.org/10.1007/JHEP02(2022)005}{\emph{JHEP} {\bf 02} (2022)
  005}, [\href{https://arxiv.org/abs/2107.10274}{{\tt 2107.10274}}].

\bibitem{Nakayama:2016knq}
Y.~Nakayama, \emph{{Bootstrap bound for conformal multi-flavor QCD on
  lattice}}, \href{http://dx.doi.org/10.1007/JHEP07(2016)038}{\emph{JHEP} {\bf
  07} (2016) 038}, [\href{https://arxiv.org/abs/1605.04052}{{\tt 1605.04052}}].

\bibitem{Iha:2016ppj}
H.~Iha, H.~Makino and H.~Suzuki, \emph{{Upper bound on the mass anomalous
  dimension in many-flavor gauge theories: a conformal bootstrap approach}},
  \href{http://dx.doi.org/10.1093/ptep/ptw046}{\emph{PTEP} {\bf 2016} (2016)
  053B03}, [\href{https://arxiv.org/abs/1603.01995}{{\tt 1603.01995}}].

\bibitem{Poland:2010wg}
D.~Poland and D.~Simmons-Duffin, \emph{{Bounds on 4D Conformal and
  Superconformal Field Theories}},
  \href{http://dx.doi.org/10.1007/JHEP05(2011)017}{\emph{JHEP} {\bf 05} (2011)
  017}, [\href{https://arxiv.org/abs/1009.2087}{{\tt 1009.2087}}].

\bibitem{Poland:2011ey}
D.~Poland, D.~Simmons-Duffin and A.~Vichi, \emph{{Carving Out the Space of 4D
  CFTs}}, \href{http://dx.doi.org/10.1007/JHEP05(2012)110}{\emph{JHEP} {\bf 05}
  (2012) 110}, [\href{https://arxiv.org/abs/1109.5176}{{\tt 1109.5176}}].

\bibitem{Berkooz:2014yda}
M.~Berkooz, R.~Yacoby and A.~Zait, \emph{{Bounds on $\mathcal{N} = 1$
  superconformal theories with global symmetries}},
  \href{http://dx.doi.org/10.1007/JHEP01(2015)132,
  10.1007/JHEP08(2014)008}{\emph{JHEP} {\bf 08} (2014) 008},
  [\href{https://arxiv.org/abs/1402.6068}{{\tt 1402.6068}}].

\bibitem{Li:2017ddj}
D.~Li, D.~Meltzer and A.~Stergiou, \emph{{Bootstrapping mixed correlators in 4D
  $ \mathcal{N} $ = 1 SCFTs}},
  \href{http://dx.doi.org/10.1007/JHEP07(2017)029}{\emph{JHEP} {\bf 07} (2017)
  029}, [\href{https://arxiv.org/abs/1702.00404}{{\tt 1702.00404}}].

\bibitem{Lin:2019vgi}
Y.-H. Lin, D.~Meltzer, S.-H. Shao and A.~Stergiou, \emph{{Bounds on Triangle
  Anomalies in (3+1)d}},
  \href{http://dx.doi.org/10.1103/PhysRevD.101.125007}{\emph{Phys. Rev. D} {\bf
  101} (2020) 125007}, [\href{https://arxiv.org/abs/1909.11676}{{\tt
  1909.11676}}].

\bibitem{Poland:2015mta}
D.~Poland and A.~Stergiou, \emph{{Exploring the Minimal 4D $\mathcal{N}=1$
  SCFT}}, \href{http://dx.doi.org/10.1007/JHEP12(2015)121}{\emph{JHEP} {\bf 12}
  (2015) 121}, [\href{https://arxiv.org/abs/1509.06368}{{\tt 1509.06368}}].

\bibitem{Beem:2014zpa}
C.~Beem, M.~Lemos, P.~Liendo, L.~Rastelli and B.~C. van Rees, \emph{{The $
  \mathcal{N}=2 $ superconformal bootstrap}},
  \href{http://dx.doi.org/10.1007/JHEP03(2016)183}{\emph{JHEP} {\bf 03} (2016)
  183}, [\href{https://arxiv.org/abs/1412.7541}{{\tt 1412.7541}}].

\bibitem{Lemos:2015awa}
M.~Lemos and P.~Liendo, \emph{{Bootstrapping $ \mathcal{N}=2 $ chiral
  correlators}}, \href{http://dx.doi.org/10.1007/JHEP01(2016)025}{\emph{JHEP}
  {\bf 01} (2016) 025}, [\href{https://arxiv.org/abs/1510.03866}{{\tt
  1510.03866}}].

\bibitem{Cornagliotto:2017snu}
M.~Cornagliotto, M.~Lemos and P.~Liendo, \emph{{Bootstrapping the $(A_1,A_2)$
  Argyres-Douglas theory}},
  \href{http://dx.doi.org/10.1007/JHEP03(2018)033}{\emph{JHEP} {\bf 03} (2018)
  033}, [\href{https://arxiv.org/abs/1711.00016}{{\tt 1711.00016}}].

\bibitem{Gimenez-Grau:2020jrx}
A.~Gimenez-Grau and P.~Liendo, \emph{{Bootstrapping Coulomb and Higgs branch
  operators}}, \href{http://dx.doi.org/10.1007/JHEP01(2021)175}{\emph{JHEP}
  {\bf 01} (2021) 175}, [\href{https://arxiv.org/abs/2006.01847}{{\tt
  2006.01847}}].

\bibitem{Bissi:2021rei}
A.~Bissi, F.~Fucito, A.~Manenti, J.~F. Morales and R.~Savelli, \emph{{OPE
  coefficients in Argyres-Douglas theories}},
  \href{https://arxiv.org/abs/2112.11899}{{\tt 2112.11899}}.

\bibitem{Beem:2013qxa}
C.~Beem, L.~Rastelli and B.~C. van Rees, \emph{{The $\mathcal{N}=4$
  Superconformal Bootstrap}},
  \href{http://dx.doi.org/10.1103/PhysRevLett.111.071601}{\emph{Phys.Rev.Lett.}
  {\bf 111} (2013) 071601}, [\href{https://arxiv.org/abs/1304.1803}{{\tt
  1304.1803}}].

\bibitem{Alday:2013opa}
L.~F. Alday and A.~Bissi, \emph{{The superconformal bootstrap for structure
  constants}}, \href{http://dx.doi.org/10.1007/JHEP09(2014)144}{\emph{JHEP}
  {\bf 09} (2014) 144}, [\href{https://arxiv.org/abs/1310.3757}{{\tt
  1310.3757}}].

\bibitem{Beem:2016wfs}
C.~Beem, L.~Rastelli and B.~C. van Rees, \emph{{More ${\mathcal N}=4$
  superconformal bootstrap}},
  \href{http://dx.doi.org/10.1103/PhysRevD.96.046014}{\emph{Phys. Rev.} {\bf
  D96} (2017) 046014}, [\href{https://arxiv.org/abs/1612.02363}{{\tt
  1612.02363}}].

\bibitem{Beem:2015aoa}
C.~Beem, M.~Lemos, L.~Rastelli and B.~C. van Rees, \emph{{The (2, 0)
  superconformal bootstrap}},
  \href{http://dx.doi.org/10.1103/PhysRevD.93.025016}{\emph{Phys. Rev.} {\bf
  D93} (2016) 025016}, [\href{https://arxiv.org/abs/1507.05637}{{\tt
  1507.05637}}].

\bibitem{Lemos:2021azv}
M.~Lemos, B.~C. van Rees and X.~Zhao, \emph{{Regge trajectories for the (2, 0)
  theories}}, \href{http://dx.doi.org/10.1007/JHEP01(2022)022}{\emph{JHEP} {\bf
  01} (2022) 022}, [\href{https://arxiv.org/abs/2105.13361}{{\tt 2105.13361}}].

\bibitem{Bissi:2020jve}
A.~Bissi, A.~Manenti and A.~Vichi, \emph{{Bootstrapping mixed correlators in $
  \mathcal{N} $ = 4 super Yang-Mills}},
  \href{http://dx.doi.org/10.1007/JHEP05(2021)111}{\emph{JHEP} {\bf 05} (2021)
  111}, [\href{https://arxiv.org/abs/2010.15126}{{\tt 2010.15126}}].

\bibitem{Alday:2021vfb}
L.~F. Alday, S.~M. Chester and T.~Hansen, \emph{{Modular invariant holographic
  correlators for $ \mathcal{N} $ = 4 SYM with general gauge group}},
  \href{http://dx.doi.org/10.1007/JHEP12(2021)159}{\emph{JHEP} {\bf 12} (2021)
  159}, [\href{https://arxiv.org/abs/2110.13106}{{\tt 2110.13106}}].

\bibitem{Chester:2021aun}
S.~M. Chester, R.~Dempsey and S.~S. Pufu, \emph{{Bootstrapping $\mathcal{N}=4$
  super-Yang-Mills on the conformal manifold}},
  \href{https://arxiv.org/abs/2111.07989}{{\tt 2111.07989}}.

\bibitem{Gaiotto:2013nva}
D.~Gaiotto, D.~Mazac and M.~F. Paulos, \emph{{Bootstrapping the 3d Ising twist
  defect}}, \href{http://dx.doi.org/10.1007/JHEP03(2014)100}{\emph{JHEP} {\bf
  03} (2014) 100}, [\href{https://arxiv.org/abs/1310.5078}{{\tt 1310.5078}}].

\bibitem{Ghosh:2021ruh}
K.~Ghosh, A.~Kaviraj and M.~F. Paulos, \emph{{Charging up the functional
  bootstrap}}, \href{http://dx.doi.org/10.1007/JHEP10(2021)116}{\emph{JHEP}
  {\bf 10} (2021) 116}, [\href{https://arxiv.org/abs/2107.00041}{{\tt
  2107.00041}}].

\bibitem{Liendo:2012hy}
P.~Liendo, L.~Rastelli and B.~C. van Rees, \emph{{The Bootstrap Program for
  Boundary CFT${}_d$}},
  \href{http://dx.doi.org/10.1007/JHEP07(2013)113}{\emph{JHEP} {\bf 1307}
  (2013) 113}, [\href{https://arxiv.org/abs/1210.4258}{{\tt 1210.4258}}].

\bibitem{Gliozzi:2015qsa}
F.~Gliozzi, P.~Liendo, M.~Meineri and A.~Rago, \emph{{Boundary and Interface
  CFTs from the Conformal Bootstrap}},
  \href{http://dx.doi.org/10.1007/JHEP05(2015)036}{\emph{JHEP} {\bf 05} (2015)
  036}, [\href{https://arxiv.org/abs/1502.07217}{{\tt 1502.07217}}].

\bibitem{Gliozzi:2016cmg}
F.~Gliozzi, \emph{{Truncatable bootstrap equations in algebraic form and
  critical surface exponents}},
  \href{http://dx.doi.org/10.1007/JHEP10(2016)037}{\emph{JHEP} {\bf 10} (2016)
  037}, [\href{https://arxiv.org/abs/1605.04175}{{\tt 1605.04175}}].

\bibitem{Padayasi:2021sik}
J.~Padayasi, A.~Krishnan, M.~A. Metlitski, I.~A. Gruzberg and M.~Meineri,
  \emph{{The extraordinary boundary transition in the 3d O(N) model via
  conformal bootstrap}},  \href{https://arxiv.org/abs/2111.03071}{{\tt
  2111.03071}}.

\bibitem{Behan:2020nsf}
C.~Behan, L.~Di~Pietro, E.~Lauria and B.~C. Van~Rees, \emph{{Bootstrapping
  boundary-localized interactions}},
  \href{http://dx.doi.org/10.1007/JHEP12(2020)182}{\emph{JHEP} {\bf 12} (2020)
  182}, [\href{https://arxiv.org/abs/2009.03336}{{\tt 2009.03336}}].

\bibitem{Behan:2021tcn}
C.~Behan, L.~Di~Pietro, E.~Lauria and B.~C. van Rees, \emph{{Bootstrapping
  boundary-localized interactions II. Minimal models at the boundary}},
  \href{http://dx.doi.org/10.1007/JHEP03(2022)146}{\emph{JHEP} {\bf 03} (2022)
  146}, [\href{https://arxiv.org/abs/2111.04747}{{\tt 2111.04747}}].

\bibitem{Collier:2021ngi}
S.~Collier, D.~Mazac and Y.~Wang, \emph{{Bootstrapping Boundaries and Branes}},
   \href{https://arxiv.org/abs/2112.00750}{{\tt 2112.00750}}.

\bibitem{Liendo:2016ymz}
P.~Liendo and C.~Meneghelli, \emph{{Bootstrap equations for $ \mathcal{N} $ = 4
  SYM with defects}},
  \href{http://dx.doi.org/10.1007/JHEP01(2017)122}{\emph{JHEP} {\bf 01} (2017)
  122}, [\href{https://arxiv.org/abs/1608.05126}{{\tt 1608.05126}}].

\bibitem{Liendo:2018ukf}
P.~Liendo, C.~Meneghelli and V.~Mitev, \emph{{Bootstrapping the half-BPS line
  defect}}, \href{http://dx.doi.org/10.1007/JHEP10(2018)077}{\emph{JHEP} {\bf
  10} (2018) 077}, [\href{https://arxiv.org/abs/1806.01862}{{\tt 1806.01862}}].

\bibitem{Gimenez-Grau:2019hez}
A.~Gimenez-Grau and P.~Liendo, \emph{{Bootstrapping line defects in
  $\mathcal{N}=2$ theories}},
  \href{http://dx.doi.org/10.1007/JHEP03(2020)121}{\emph{JHEP} {\bf 03} (2020)
  121}, [\href{https://arxiv.org/abs/1907.04345}{{\tt 1907.04345}}].

\bibitem{Cavaglia:2021bnz}
A.~Cavagli\`a, N.~Gromov, J.~Julius and M.~Preti, \emph{{Integrability and
  conformal bootstrap: One dimensional defect conformal field theory}},
  \href{http://dx.doi.org/10.1103/PhysRevD.105.L021902}{\emph{Phys. Rev. D}
  {\bf 105} (2022) L021902}, [\href{https://arxiv.org/abs/2107.08510}{{\tt
  2107.08510}}].

\bibitem{Cavaglia:2022qpg}
A.~Cavagli\`a, N.~Gromov, J.~Julius and M.~Preti, \emph{{Bootstrability in
  Defect CFT: Integrated Correlators and Sharper Bounds}},
  \href{https://arxiv.org/abs/2203.09556}{{\tt 2203.09556}}.

\bibitem{cohnelkies}
H.~Cohn and N.~Elkies, \emph{New upper bounds on sphere packings i},
  \href{http://dx.doi.org/10.4007/annals.2003.157.689}{\emph{Annals of
  Mathematics} {\bf 157} (Mar, 2003) 689–714}.

\bibitem{Kos:2013tga}
F.~Kos, D.~Poland and D.~Simmons-Duffin, \emph{{Bootstrapping the $O(N)$ vector
  models}}, \href{http://dx.doi.org/10.1007/JHEP06(2014)091}{\emph{JHEP} {\bf
  1406} (2014) 091}, [\href{https://arxiv.org/abs/1307.6856}{{\tt 1307.6856}}].

\bibitem{Simmons-Duffin:2015qma}
D.~Simmons-Duffin, \emph{{A Semidefinite Program Solver for the Conformal
  Bootstrap}}, \href{http://dx.doi.org/10.1007/JHEP06(2015)174}{\emph{JHEP}
  {\bf 06} (2015) 174}, [\href{https://arxiv.org/abs/1502.02033}{{\tt
  1502.02033}}].

\bibitem{poulson2013elemental}
J.~Poulson, B.~Marker, R.~A. Van~de Geijn, J.~R. Hammond and N.~A. Romero,
  \emph{Elemental: A new framework for distributed memory dense matrix
  computations}, {\emph{ACM Transactions on Mathematical Software (TOMS)} {\bf
  39} (2013) 13}.

\bibitem{petschow2013high}
M.~Petschow, E.~Peise and P.~Bientinesi, \emph{High-performance solvers for
  dense hermitian eigenproblems}, {\emph{SIAM Journal on Scientific Computing}
  {\bf 35} (2013) C1--C22}.

\bibitem{Landry:2019qug}
W.~Landry and D.~Simmons-Duffin, \emph{{Scaling the semidefinite program solver
  SDPB}},  \href{https://arxiv.org/abs/1909.09745}{{\tt 1909.09745}}.

\bibitem{sdpb}
W.~Landry and D.~Simmons-Duffin. \url{https://github.com/davidsd/sdpb}.

\bibitem{Paulos:2014vya}
M.~F. Paulos, \emph{{JuliBootS: a hands-on guide to the conformal bootstrap}},
  \href{https://arxiv.org/abs/1412.4127}{{\tt 1412.4127}}.

\bibitem{Paulos:2019fkw}
M.~F. Paulos and B.~Zan, \emph{{A functional approach to the numerical
  conformal bootstrap}},
  \href{http://dx.doi.org/10.1007/JHEP09(2020)006}{\emph{JHEP} {\bf 09} (2020)
  006}, [\href{https://arxiv.org/abs/1904.03193}{{\tt 1904.03193}}].

\bibitem{Mazac:2019shk}
D.~Maz\'a\v{c}, L.~Rastelli and X.~Zhou, \emph{{A basis of analytic functionals
  for CFTs in general dimension}},
  \href{http://dx.doi.org/10.1007/JHEP08(2021)140}{\emph{JHEP} {\bf 08} (2021)
  140}, [\href{https://arxiv.org/abs/1910.12855}{{\tt 1910.12855}}].

\bibitem{Penedones:2019tng}
J.~Penedones, J.~A. Silva and A.~Zhiboedov, \emph{{Nonperturbative Mellin
  Amplitudes: Existence, Properties, Applications}},
  \href{http://dx.doi.org/10.1007/JHEP08(2020)031}{\emph{JHEP} {\bf 08} (2020)
  031}, [\href{https://arxiv.org/abs/1912.11100}{{\tt 1912.11100}}].

\bibitem{Kologlu:2019bco}
M.~Kologlu, P.~Kravchuk, D.~Simmons-Duffin and A.~Zhiboedov, \emph{{Shocks,
  Superconvergence, and a Stringy Equivalence Principle}},
  \href{http://dx.doi.org/10.1007/JHEP11(2020)096}{\emph{JHEP} {\bf 11} (2020)
  096}, [\href{https://arxiv.org/abs/1904.05905}{{\tt 1904.05905}}].

\bibitem{Carmi:2019cub}
D.~Carmi and S.~Caron-Huot, \emph{{A Conformal Dispersion Relation:
  Correlations from Absorption}},
  \href{http://dx.doi.org/10.1007/JHEP09(2020)009}{\emph{JHEP} {\bf 09} (2020)
  009}, [\href{https://arxiv.org/abs/1910.12123}{{\tt 1910.12123}}].

\bibitem{Caron-Huot:2020adz}
S.~Caron-Huot, D.~Mazac, L.~Rastelli and D.~Simmons-Duffin, \emph{{Dispersive
  CFT Sum Rules}}, \href{http://dx.doi.org/10.1007/JHEP05(2021)243}{\emph{JHEP}
  {\bf 05} (2021) 243}, [\href{https://arxiv.org/abs/2008.04931}{{\tt
  2008.04931}}].

\bibitem{Afkhami-Jeddi:2019zci}
N.~Afkhami-Jeddi, T.~Hartman and A.~Tajdini, \emph{{Fast Conformal Bootstrap
  and Constraints on 3d Gravity}},
  \href{http://dx.doi.org/10.1007/JHEP05(2019)087}{\emph{JHEP} {\bf 05} (2019)
  087}, [\href{https://arxiv.org/abs/1903.06272}{{\tt 1903.06272}}].

\bibitem{Hellerman:2009bu}
S.~Hellerman, \emph{{A Universal Inequality for CFT and Quantum Gravity}},
  \href{http://dx.doi.org/10.1007/JHEP08(2011)130}{\emph{JHEP} {\bf 08} (2011)
  130}, [\href{https://arxiv.org/abs/0902.2790}{{\tt 0902.2790}}].

\bibitem{Collier:2016cls}
S.~Collier, Y.-H. Lin and X.~Yin, \emph{{Modular Bootstrap Revisited}},
  \href{http://dx.doi.org/10.1007/JHEP09(2018)061}{\emph{JHEP} {\bf 09} (2018)
  061}, [\href{https://arxiv.org/abs/1608.06241}{{\tt 1608.06241}}].

\bibitem{Arkani-Hamed:2018ign}
N.~Arkani-Hamed, Y.-T. Huang and S.-H. Shao, \emph{{On the Positive Geometry of
  Conformal Field Theory}},
  \href{http://dx.doi.org/10.1007/JHEP06(2019)124}{\emph{JHEP} {\bf 06} (2019)
  124}, [\href{https://arxiv.org/abs/1812.07739}{{\tt 1812.07739}}].

\bibitem{El-Showk:2012vjm}
S.~El-Showk and M.~F. Paulos, \emph{{Bootstrapping Conformal Field Theories
  with the Extremal Functional Method}},
  \href{http://dx.doi.org/10.1103/PhysRevLett.111.241601}{\emph{Phys. Rev.
  Lett.} {\bf 111} (2013) 241601}, [\href{https://arxiv.org/abs/1211.2810}{{\tt
  1211.2810}}].

\bibitem{El-Showk:2016mxr}
S.~El-Showk and M.~F. Paulos, \emph{{Extremal bootstrapping: go with the
  flow}}, \href{http://dx.doi.org/10.1007/JHEP03(2018)148}{\emph{JHEP} {\bf 03}
  (2018) 148}, [\href{https://arxiv.org/abs/1605.08087}{{\tt 1605.08087}}].

\bibitem{Liu:2020tpf}
J.~Liu, D.~Meltzer, D.~Poland and D.~Simmons-Duffin, \emph{{The Lorentzian
  inversion formula and the spectrum of the 3d O(2) CFT}},
  \href{http://dx.doi.org/10.1007/JHEP09(2020)115}{\emph{JHEP} {\bf 09} (2020)
  115}, [\href{https://arxiv.org/abs/2007.07914}{{\tt 2007.07914}}].

\bibitem{Gliozzi:2013ysa}
F.~Gliozzi, \emph{{More constraining conformal bootstrap}},
  \href{http://dx.doi.org/10.1103/PhysRevLett.111.161602}{\emph{Phys. Rev.
  Lett.} {\bf 111} (2013) 161602}, [\href{https://arxiv.org/abs/1307.3111}{{\tt
  1307.3111}}].

\bibitem{Afkhami-Jeddi:2021iuw}
N.~Afkhami-Jeddi, \emph{{Conformal Bootstrap Deformations}},
  \href{https://arxiv.org/abs/2111.01799}{{\tt 2111.01799}}.

\bibitem{Gliozzi:2014jsa}
F.~Gliozzi and A.~Rago, \emph{{Critical exponents of the 3d Ising and related
  models from Conformal Bootstrap}},
  \href{http://dx.doi.org/10.1007/JHEP10(2014)042}{\emph{JHEP} {\bf 1410}
  (2014) 42}, [\href{https://arxiv.org/abs/1403.6003}{{\tt 1403.6003}}].

\bibitem{Iliesiu:2018fao}
L.~Iliesiu, M.~Kolo\u{g}lu, R.~Mahajan, E.~Perlmutter and D.~Simmons-Duffin,
  \emph{{The Conformal Bootstrap at Finite Temperature}},
  \href{http://dx.doi.org/10.1007/JHEP10(2018)070}{\emph{JHEP} {\bf 10} (2018)
  070}, [\href{https://arxiv.org/abs/1802.10266}{{\tt 1802.10266}}].

\bibitem{Iliesiu:2018zlz}
L.~Iliesiu, M.~Kolo\u{g}lu and D.~Simmons-Duffin, \emph{{Bootstrapping the 3d
  Ising model at finite temperature}},
  \href{http://dx.doi.org/10.1007/JHEP12(2019)072}{\emph{JHEP} {\bf 12} (2019)
  072}, [\href{https://arxiv.org/abs/1811.05451}{{\tt 1811.05451}}].

\bibitem{Reehorst:2021ykw}
M.~Reehorst, S.~Rychkov, D.~Simmons-Duffin, B.~Sirois, N.~Su and B.~van Rees,
  \emph{{Navigator Function for the Conformal Bootstrap}},
  \href{http://dx.doi.org/10.21468/SciPostPhys.11.3.072}{\emph{SciPost Phys.}
  {\bf 11} (2021) 072}, [\href{https://arxiv.org/abs/2104.09518}{{\tt
  2104.09518}}].

\bibitem{Fitzpatrick:2012yx}
A.~L. Fitzpatrick, J.~Kaplan, D.~Poland and D.~Simmons-Duffin, \emph{{The
  Analytic Bootstrap and AdS Superhorizon Locality}},
  \href{http://dx.doi.org/10.1007/JHEP12(2013)004}{\emph{JHEP} {\bf 1312}
  (2013) 004}, [\href{https://arxiv.org/abs/1212.3616}{{\tt 1212.3616}}].

\bibitem{Komargodski:2012ek}
Z.~Komargodski and A.~Zhiboedov, \emph{{Convexity and Liberation at Large
  Spin}}, \href{http://dx.doi.org/10.1007/JHEP11(2013)140}{\emph{JHEP} {\bf
  1311} (2013) 140}, [\href{https://arxiv.org/abs/1212.4103}{{\tt 1212.4103}}].

\bibitem{Lin:2020mme}
H.~W. Lin, \emph{{Bootstraps to strings: solving random matrix models with
  positivite}}, \href{http://dx.doi.org/10.1007/JHEP06(2020)090}{\emph{JHEP}
  {\bf 06} (2020) 090}, [\href{https://arxiv.org/abs/2002.08387}{{\tt
  2002.08387}}].

\bibitem{Han:2020bkb}
X.~Han, S.~A. Hartnoll and J.~Kruthoff, \emph{{Bootstrapping Matrix Quantum
  Mechanics}},
  \href{http://dx.doi.org/10.1103/PhysRevLett.125.041601}{\emph{Phys. Rev.
  Lett.} {\bf 125} (2020) 041601},
  [\href{https://arxiv.org/abs/2004.10212}{{\tt 2004.10212}}].

\bibitem{Kazakov:2021lel}
V.~Kazakov and Z.~Zheng, \emph{{Analytic and Numerical Bootstrap for One-Matrix
  Model and ''Unsolvable'' Two-Matrix Model}},
  \href{https://arxiv.org/abs/2108.04830}{{\tt 2108.04830}}.

\bibitem{Bonifacio:2020xoc}
J.~Bonifacio and K.~Hinterbichler, \emph{{Bootstrap Bounds on Closed Einstein
  Manifolds}}, \href{http://dx.doi.org/10.1007/JHEP10(2020)069}{\emph{JHEP}
  {\bf 10} (2020) 069}, [\href{https://arxiv.org/abs/2007.10337}{{\tt
  2007.10337}}].

\bibitem{Kravchuk:2021akc}
P.~Kravchuk, D.~Mazac and S.~Pal, \emph{{Automorphic Spectra and the Conformal
  Bootstrap}},  \href{https://arxiv.org/abs/2111.12716}{{\tt 2111.12716}}.

\bibitem{Bonifacio:2021aqf}
J.~Bonifacio, \emph{{Bootstrapping closed hyperbolic surfaces}},
  \href{http://dx.doi.org/10.1007/JHEP03(2022)093}{\emph{JHEP} {\bf 03} (2022)
  093}, [\href{https://arxiv.org/abs/2111.13215}{{\tt 2111.13215}}].

\bibitem{Hartman:2019pcd}
T.~Hartman, D.~Maz\'a\v{c} and L.~Rastelli, \emph{{Sphere Packing and Quantum
  Gravity}}, \href{http://dx.doi.org/10.1007/JHEP12(2019)048}{\emph{JHEP} {\bf
  12} (2019) 048}, [\href{https://arxiv.org/abs/1905.01319}{{\tt 1905.01319}}].

\bibitem{Afkhami-Jeddi:2020hde}
N.~Afkhami-Jeddi, H.~Cohn, T.~Hartman, D.~de~Laat and A.~Tajdini,
  \emph{{High-dimensional sphere packing and the modular bootstrap}},
  \href{http://dx.doi.org/10.1007/JHEP12(2020)066}{\emph{JHEP} {\bf 12} (2020)
  066}, [\href{https://arxiv.org/abs/2006.02560}{{\tt 2006.02560}}].

\bibitem{Behan:2016dtz}
C.~Behan, \emph{{PyCFTBoot: A flexible interface for the conformal bootstrap}},
  \href{http://dx.doi.org/10.4208/cicp.OA-2016-0107}{\emph{Commun. Comput.
  Phys.} {\bf 22} (2017) 1--38}, [\href{https://arxiv.org/abs/1602.02810}{{\tt
  1602.02810}}].

\bibitem{cboot}
T.~Ohtsuki. \url{https://github.com/tohtsky/cboot}.

\bibitem{simpleboot}
N.~Su. \url{https://gitlab.com/bootstrapcollaboration/simpleboot}.

\bibitem{hyperion-bootstrap}
D.~Simmons-Duffin et~al. \url{https://gitlab.com/davidsd/hyperion-bootstrap}.

\bibitem{scalarblocks}
W.~Landry et~al. \url{https://gitlab.com/bootstrapcollaboration/scalar_blocks}.

\bibitem{Cuomo:2017wme}
G.~F. Cuomo, D.~Karateev and P.~Kravchuk, \emph{{General Bootstrap Equations in
  4D CFTs}}, \href{http://dx.doi.org/10.1007/JHEP01(2018)130}{\emph{JHEP} {\bf
  01} (2018) 130}, [\href{https://arxiv.org/abs/1705.05401}{{\tt 1705.05401}}].

\bibitem{virasoro}
C.~Hussong. \url{https://github.com/chussong/virasoro}.

\bibitem{quadraticnet}
D.~Simmons-Duffin et~al. \url{https://gitlab.com/davidsd/quadratic-net/}.

\bibitem{tiptop}
W.~Landry, D.~Simmons-Duffin et~al.
  \url{https://gitlab.com/bootstrapcollaboration/tiptop}.

\bibitem{Komargodski:2016auf}
Z.~Komargodski and D.~Simmons-Duffin, \emph{{The Random-Bond Ising Model in
  2.01 and 3 Dimensions}},
  \href{http://dx.doi.org/10.1088/1751-8121/aa6087}{\emph{J. Phys. A} {\bf 50}
  (2017) 154001}, [\href{https://arxiv.org/abs/1603.04444}{{\tt 1603.04444}}].

\end{thebibliography}\endgroup

\end{document}